**TITLE**
**Recurrent Network Models Of Sequence Generation And Memory**

**AUTHORS**
Kanaka Rajan, Christopher D Harvey and David W Tank

**AFFILIATIONS**
Joseph Henry Laboratories of Physics and Lewis–Sigler Institute for Integrative Genomics, Princeton University, Princeton, NJ.
Department of Neurobiology, Harvard Medical School, Boston, MA.
Department of Molecular Biology and Princeton Neuroscience Institute, Princeton University, Princeton, NJ.

**CORRESPONDENCE**
KR: *krajan@princeton.edu*
CDH: harvey@hms.harvard.edu
DWT: dwtank@princeton.edu

**SUMMARY**

The sequential activation of neurons is a common feature of network activity during a variety of behaviors and has been proposed as a mechanism for cortical computation, including short term memory. Previous modeling approaches for sequences and memory networks have emphasized highly specialized architectures in which a principled mechanism is pre-wired into the connectivity of the network. Here, we demonstrate that starting from random synaptic connectivity and allowing a small fraction of connections to undergo modification, a largely disordered recurrent network can produce sequences and short-term memory. We use this process, which we call Partial In-Network training (PINning), to model and match data from cellular-resolution imaging of neural activity in the mouse posterior parietal cortex (PPC) during a memory-guided two-alternative forced choice task in a virtual environment [Harvey, Coen & Tank, 2012]. In the model, as in the PPC data, individual neurons exhibit transient activations that are staggered relative to one another in time to form sequences spanning the duration of the task, and different sequences are activated on trials with different cues and choices. Analysis of the connectivity matrices of the minimally structured model networks revealed that the time-ordered neural activity is produced by the cooperation between recurrent synaptic interactions and external inputs, rather than feedforward connections, or the asymmetric connections of ring attractor models for sequences. In addition, our model showed that sequential activation across a population of neurons is an efficient mechanism for implementing short-term memory with comparable memory capabilities to previously proposed fixed point mechanisms. Together our results develop a new modeling framework based on generic, minimally modified networks and suggest that neural activity sequences may emerge through learning from largely unstructured network architectures.

**INTRODUCTION**

Sequential firing has emerged as a prominent motif of population activity in several experiments involving temporally structured behaviors, such as short-term memory and decision making. Neural sequences have been observed in many brain regions including the cortex [Luczak et al, 2007; Schwartz & Moran, 1999; Andersen et al, 2004; Pulvermutter & Shtyrov, 2009; Buonomano, 2003; Ikegaya et al, 2004; Tang et al, 2008; Seidemann et al, 1996; Fujisawa et al, 2008; Crowe et al, 2010; Harvey, Coen & Tank, 2012], hippocampus [Nadasdy et al, 1999; Louie & Wilson, 2001; Pastalkova et al, 2008; Davidson, Kloosterman & Wilson, 2009], basal ganglia [Barnes et al, 2005; Jin, Fuji & Graybiel, 2009], cerebellum [Mauk & Buonomano, 2004], and area HVC of the songbird [Hahnloser, Kozhevnikov & Fee, 2002; Kozhevnikov & Fee, 2007]. In all these cases, the observed sequences span a wide range of time durations, but individual neurons fire transiently only during a small portion of the full sequence. The ubiquity of neural sequences in different brain regions suggests that they are of widespread functional use, and that they may be produced by general circuit-level mechanisms.

Broadly speaking, sequences can be produced by highly structured neural circuits or by more generic circuits adapted through the learning of a specific task. Highly structured circuits of this type have a long history [Kleinfeld & Sompolinsky, 1989; Goldman, 2009], for example, as synfire chain models [Hertz & Prugel-Bennett,1996; Levy et al, 2001; Hermann, Hertz & Prugel-Bennet, 1995; Fiete et al, 2010], in which excitation flows unidirectionally from one active neuron to the next along a chain of connected neurons, or as ring attractor models [Yishai, Bar-Or & Sompolinsky, 1995; Zhang, 1996], in which increased ("central") excitation between nearby neurons surrounded by long-range inhibition and asymmetric connectivity are responsible for time-ordered neural activity. Constructing these models typically involves implementing a task-specific mechanism (for producing sequences, for instance) into the synaptic connectivity, producing highly specialized networks. Neural circuits are highly adaptive and involved in a wide variety of tasks, and furthermore, sequential neural activity often emerges through the learning of a specific task and retains significant variability. It is therefore unlikely for highly structured approaches to produce models with flexible circuitry or to generate dynamics with the temporal complexity needed to make a connection with experimental recordings of neural sequences.

In contrast, random networks of model neurons interconnected with excitatory and inhibitory connections in a balanced state [Sompolinsky, Crisanti & Sommers, 1988], rather than being specifically designed for one single task, have been modified by training to perform a variety of tasks [Buonomano & Merzenich, 1995; Buonomano, 2005; Williams & Zipser, 1989; Pearlmutter, 1989; Jaeger & Haas, 2004; Maass, Joshi & Sontag, 2007; Sussillo & Abbott, 2009; Maass, Natschlager & Markram, 2002; Jaeger, 2003]. Here, we built on these lines of research and asked whether a general implementation using relatively unstructured random networks could create sequential neural dynamics resembling experimental data. We used data from sequences observed in the posterior parietal cortex (PPC) of mice trained to perform a two-alternative forced-choice (2AFC) task in a virtual reality environment [Harvey, Coen & Tank, 2012], and also constructed models that extrapolated beyond these experimental data.

To address how much network structure is required to support sequences like those observed in the neural recordings, we introduced a novel modeling framework called **Partial In-Network Training** or PINning. In this scheme, any desired fraction of the initially random connections within the networks we construct can be modified by a synaptic change algorithm, enabling us to explore the full range of networks between completely random and fully structured. Using network models constructed by PINning, we first demonstrated that sequences resembling the PPC data are most consistent with minimally structured neural circuitry, with small amounts of structured connectivity that supports sequential activity patterns and a much larger fraction of unstructured connections. Next, we investigated the circuit-level mechanism of sequence generation in largely random networks containing some learned structure. Finally, we determined the role modeled sequences play in short-term memory, for instance, by storing information during the delay period about whether a left or right turn was indicated early in the 2AFC task [Harvey, Coen & Tank, 2012]. Going beyond models meant to reproduce the experimental data, we also analyzed multiple sequences initiated by different sensory cues and computed the capacity of this form of short-term memory.

**RESULTS**

1. **Sequences from highly structured or random networks do not match PPC data**

Networks of rate-based model neurons, in which the outputs of individual neurons are characterized by firing rates and units are interconnected through excitatory and inhibitory synapses of various strengths (Experimental Procedures 1), form the basis of our studies. To interpret the outputs of the rate networks we construct in terms of experimental data, we extract firing rates from the calcium fluorescence signals recorded in the PPC using two complementary deconvolution methods (Figures 1C and D show calcium data from [Harvey, Coen & Tank, 2012] and the rates extracted from these data, respectively; see also Supplemental Figure 1 and Experimental Procedures 3). We define two measures to compare the rates from the model to the rates extracted from data. The first, *bVar*, measures the stereotypy of the data or the network output by quantifying the variance that is explained by the translation along the sequence of an activity profile with an invariant shape (Figure 1G and Experimental Procedures 6). The second metric, *pVar*, quantifies the percent variance of the experimental data from the PPC [Harvey, Coen & Tank, 2012] that is captured by the outputs of the different networks we build and is therefore useful for tracking their performance as a function of network parameters (Figure 1H and Experimental Procedures 7). *bVar* and *pVar* are used throughout this paper (for the results in Figures 1–6).

Many models have suggested that highly structured synaptic connectivity, i.e., containing ring-like or chain-like interactions, in networks is responsible for neural sequences [Yishai, Bar-Or & Sompolinsky, 1995; Zhang, 1996; Hertz & Prugel-Bennett,1996; Levy et al, 2001; Hermann, Hertz & Prugel-Bennet, 1995; Fiete et al, 2010]. We therefore first asked how much of the variance in long-duration neural sequences seen experimentally, for instance, in the PPC [Harvey, Coen & Tank, 2012], was consistent with a bump of activity moving across the network, as expected from such highly structured connectivity. To quantify this, we used *bVar*. The stereotypy of PPC sequences was found to be quite small – *bVar* = 40% for Figures 1D and 2A, which were averaged over hundreds of trials and pooled across different animals; in fact, *bVar* was lower in both single trial data (10–15% for data in Supplemental Figure 13 from [Harvey, Coen & Tank, 2012]) and trial-averaged data from a single mouse (15% for the data in Figure 2c from [Harvey, Coen & Tank, 2012]). The relatively small fraction of the variance explained by a moving bump (low *bVar*), combined with a weak relationship between the activity pattern of a neuron and anatomical location in the PPC data (Figure 5d in [Harvey, Coen & Tank, 2012]), motivated us to consider network architectures with disordered connectivity composed of a balanced set of excitatory and inhibitory weights drawn independently from a random distribution (Experimental Procedures 1).

In random network models, when excitation and inhibition are balanced on average, the ongoing dynamics have been shown to be chaotic [Sompolinsky, Crisanti & Sommers, 1988]; however, the presence of external stimuli can channel the ongoing dynamics in random network models by suppressing their chaos [Molgedey, Schuchhardt & Schuster, 1992; Bertschinger & Natschlager, 2004; Rajan, Abbott & Sompolinsky, 2010; Rajan, Abbott & Sompolinsky, 2011]. Furthermore, in experiments, strong inputs have been shown to reduce the Fano factor and trial-to-trial variability associated with spontaneous activity [Churchland et al, 2010; White, Abbott & Fiser, 2011]. Thus we asked whether PPC-like sequences could be constructed either from the spontaneous activity or the input-driven dynamics of such random networks. To test this, we first simulated a random network of firing rate-based model neurons operating in a chaotic regime (schematized in Figure 1A, described in Experimental Procedures 1, $N = 437$, network size chosen to match the size of the dataset under consideration). The individual firing rates were normalized by the maximum over the duration of each trial (10.5s here, consistent with Figure 1C) and sorted in ascending order of their time of center-of-mass ($t_{COM}$), matching the procedures applied to the real data [Harvey, Coen & Tank, 2012]. Although the resulting ordered chaotic spontaneous activity was sequential (not shown, but similar to Figure 1B), the level of extra-sequential background activity (*bVar* = 5 $\pm$ 2% and *pVar* = 0.15 + 0.1%) was higher than data. Sparsifying this background activity by increasing the threshold of the sigmoidal activation function increased *bVar* to a maximum of 22%, still considerably smaller than the data value of 40% (Supplemental Figure 8).

Next, we introduced external inputs to the random network that were time-varying to represent the effects of the visual stimuli in the virtual environment (a few example inputs are shown in the right panel of Figure 1A, see also Experimental Procedures 2). Another sequence was obtained by normalizing and sorting the firing rates from this input-driven random network (as with the spontaneous activity and with the data [Harvey, Coen & Tank, 2012]), but this too did not match the PPC data, *bVar* = 10 $\pm$ 2% and *pVar* = 0.2 + 0.1% (Figure 1B, see also left panel of Figure 1H and Supplemental Figure 8).

External inputs and completely disordered connectivity are insufficient to evoke sequences resembling the data (Figure 1B, see also Supplemental Figure 3), which are more structured and temporally constrained (Figures 1C–D and 2A, compared to Figure 1B). Therefore, sequences like those observed during timing and memory experiments [Harvey, Coen & Tank, 2012] are unlikely to be an inherent property of completely random networks. Furthermore, since real neural sequences arise during the learning of various experimental tasks, we asked whether initially disordered networks could also be modified by training to produce realistic sequences.

**2. Temporally Constrained Neural Sequences Emerge With Synaptic Modification**

To construct networks that match the activity seen in the PPC, we developed a training scheme, Partial In-Network Training (PINning), in which different sized subsets of synapses were modified (Experimental Procedures 4). In our synaptic modification scheme, the inputs to individual model neurons in the network were compared directly with target functions or templates derived from real experimental data [Fisher et al, 2013], on both left and right correct-choice outcome trials from 437 trial averaged neurons, pooled across 6 mice, during a 2AFC task (Figure 2A and Figure 5D). During training, the internal synaptic weights in the connectivity matrix of the recurrent network were modified using a variant of the recursive least-squares (RLS) or or first-order reduced and controlled error (FORCE) learning rule [Haykins, 2002; Sussillo & Abbott, 2009] until the network rates matched the target functions (Experimental Procedures 4). Crucially, the learning rule was applied only to all the synapses of a randomly selected and often small fraction of neurons. Only a fraction *p* ($pN^2 << N^2$) of the total number of synapses in the network were modified (plastic synapses are depicted in orange in Figures 2B, 5A and 6A). In particular, while every neuron in the network had a target function, only the outgoing synapses from a subset of neurons (i.e., only the outgoing synaptic weights from $pN$ chosen neurons) were subject to the learning rule (schematized in Figure 2B, see also Experimental Procedures 4). The remaining elements of the synaptic matrix remained unmodified and in their randomly initialized state (random synapses are depicted in gray in Figures 1A, 2B, 5A and 6A). The PINning method for building networks therefore provides a way to span the entire spectrum of possible neural architectures – from disordered networks with random connections ($p$ = 0 for Figures 1A and B), through networks with partially structured connectivity, in which only a small subset $p$ of connections are task-specifically wired ($p$ < 25% for the examples in Figures 2, 5 and 6), while the majority of connections remain disordered, to networks containing entirely trained connections ($p$ = 100%, Figure 3D–F).

We first applied PINning to a network with templates obtained from a single PPC-like sequence (Figure 2A–C) using a range of values for the fraction of plastic synapses, $p$. Modification of only a small percentage ($p$ = 12%) of the synaptic connections in an initially disordered network (Figure 1F) was sufficient for its sequential outputs to become more temporally constrained ($bVar$ = 40%) and to match the PPC data with high fidelity ($pVar$ = 85%, Figure 2C, see also Supplemental Data 12 for results of a cross-validation analysis). Although the networks shown here are typically as large as the size of the experimental dataset ($N$ = 437 for Figures 1–2 and 569 for Figure 5), our results are consistent both for larger $N$ networks and for networks in which non-targeted neurons are included to simulate the effect of unobserved but active neurons present in the experimental data (Supplemental Figure 7). The dependence of $pVar$ on $p$ is shown in Figure 2D.

Figure 2D quantifies the amount of structure required for generating sequences in terms of the relative fraction of synapses modified from their initially random values, but what is the overall magnitude of synaptic change required to produce these sequences? As shown in Figure 2E, we found that although the individual synapses changed more in sparsely PINned (small $p$) networks, the total amount of change across the synaptic connectivity matrix was smaller. We return to other implications of this issue in Section 3.

To uncover how the sequential dynamics are distributed across the population of active neurons in PINned networks, we used principal component analysis (PCA) (see for example, [Rajan, Abbott & Sompolinsky, 2011; Sussillo, 2014], and references therein). For an untrained random network (here, with $N$ = 437 and $p$ = 0) operating in a spontaneously active regime (Experimental Procedures 1), the top 38 principal components accounted for 95% of the total variance (therefore, the effective dimensionality, $Q_{eff}$ = 38) (gray circles in Figure 2F). In comparison, $Q_{eff}$ of the data in Figure 1D is 24. In the network with $p$ = 12% with outputs matching the PPC data, the dimensionality was lower ($Q_{eff}$ = 14) (orange circles in Figure 2F). $Q_{eff}$ asymptoted around 12 dimensions for higher values of $p$ (inset of Figure 2F). The circuit dynamics are higher dimensional for the output of the PINned network than for a sinusoidal bump attractor but lower than for the data.

### 3. Circuit Mechanism For Sequential Activation Through PINning

To develop a simplified prototype for further investigation of the mechanisms of sequential activation, we created a synthetic sequence "idealized" from data, lasting as long as the duration of the PPC sequence (10.5 seconds, Figure 2A). We first generated a Gaussian curve, *f(t)* (green curve in the inset of Figure 1F and in the left panel of Supplemental Figure 4A) that best fit the average extracted from the $t_{COM}$-aligned PPC data ($R_{ave}$, red curve in the inset of Figure 1F and in the left panel of Supplemental Figure 4A). The curve *f(t)* was then translated uniformly over a 10.5s time period to derive a set of target functions ($N$ = 500, right panel of Supplemental Figure 4A, *bVar* = 100%). Increasing fractions of the initially random connectivity matrix, *p*, were trained using PINning with the target functions of the idealized sequence. As before, *pVar* increases as a function of *p*, reaches *pVar* = 92% at *p* = 8% (highlighted by a red circle in Supplemental Figure 4B), asymptoting at $p \simeq$ 10%. The plasticity required for producing the idealized sequence was therefore smaller than the *p* = 12% required for the PPC-like sequence (Figure 2D), due to the lack of the idiosyncrasies present in the experimental data, for example, irregularities in the temporal ordering of the individual transients and background activity away from the sequence.

Two features are critical for the production of sequential activity in a neural circuit. The first is the formation of a subpopulation of active neurons ("bump"), maintained by excitation between co-active neurons and restricted by inhibition. The second is an asymmetry in the synaptic inputs from neurons ahead in the sequence and those behind, needed to make the bump move. We therefore looked for these two features in PINned networks by examining both the structure of their connectivity matrices and the synaptic currents into individual neurons.

In a classic moving ring/bump attractor network, the synaptic connectivity is only a function of the "distance" between pairs of network neurons in the sequence, $i - j$, assuming the neurons are labelled in order of their appearance in the sequence (for our analysis of the connectivity in PINned networks, we follow the same approach and order the neurons in a similar manner). Furthermore, the connectivity that sustains, constrains and moves the bump is all contained in the connectivity matrix, which is localized in $|i - j|$ and asymmetric. We looked for similar structure in the trained network models.

We considered 3 connectivity matrices interconnecting a population of model neurons, $N$ = 500 – before PINning (the randomly initialized matrix denoted by $J_{Rand}$), after sparse PINning ($J_{PINned, 8\%}$), and after full PINning ($J_{PINned, 100\%}$, built as a useful comparison). To analyze how the synaptic strengths in these 3 matrices varied with $i - j$, we first computed the means and the standard deviations of the diagonals, and the means and the standard deviations of successive off-diagonal "bands" moving away from the diagonals, i.e., $i - j$ = constant (Experimental Procedures 11). These band-averages and the fluctuations around them were plotted as a function of the interneuronal distance, $i - j$ (Figures 3B and 3E). Next, we generated "synthetic" interaction matrices, in which all the elements along each diagonal were replaced by their band-averages and fluctuations, respectively. Finally, these synthetic matrices were used in networks of rate-based model neurons driven by the same external inputs as the original PINned networks (Experimental Procedures 2).

In the $p$ = 100% model, the band-averages of $J_{PINned, 100\%}$ formed a localized and asymmetric profile (orange circles in Figure 3E) and produce a moving "Gaussian bump" of activity (right panel of Figure 3G) that is qualitatively similar to moving ring attractor dynamics [Yishai, Bar-Or & Sompolinsky, 1995; Zhang, 1996]. On the other hand, the band-averages for $J_{PINned, 8\%}$ (orange circles in Figure 3B) exhibited a localized zone of excitation for small values of $i - j$ that was symmetric, a significant inhibitory self-interactive or autaptic feature at $i - j$ = 0, and relatively diffuse flanking inhibition for larger values of $i - j$. This is reminiscent of the features expected in the interaction matrices of stationary bump models [Yishai, Bar-Or & Sompolinsky, 1995]. Furthermore, neither the band-averages of $J_{PINned, 8\%}$ by themselves (shown in Figure 3C) nor the fluctuations by themselves (not shown), were sufficient to produce moving sequences similar to the output of a network containing the full matrix $J_{PINned, 8\%}$. Instead, the outputs of the synthetic networks built from the components of the band-averaged $J_{PINned, 8\%}$ were stationary bumps (right panel of Figure 3C). In this case, what causes the bump to move?

To address this, we considered the fluctuations around the band averages of the sparsely PINned connectivity matrices, $J_{PINned, 8\%}$. As expected (from the color-bars in Figures 3A, the lines in Figures 3B, and Figure 3F), the fluctuations around the band-averages of $J_{PINned, 8\%}$ (red lines in Figure 3B) were much larger and more structured than those of $J_{Rand}$ (small blue lines in Figure 3B). To uncover the mechanistic role of these fluctuations, we examined the input to each neuron produced by the sum of the fluctuations of $J_{PINned, 8\%}$ around the band-averages and the external input. We realigned the sums of the fluctuations and the external inputs for all the neurons in the network by their $t_{COM}$ and then averaged over neurons (see Experimental Procedures 6 for an example of a similar procedure). This yielded an aligned population average (bottom right panel in Figure 4) that clearly revealed the asymmetry responsible for the movement of the bump across the network. Therefore, in the presence of external inputs that are constantly changing in time, the mean synaptic interactions do not have to be asymmetric, as we observed for $J_{PINned, 8\%}$ (Figure 3B). Instead, the variations in the fluctuations of $J_{PINned, 8\%}$ (i.e., after the mean has been subtracted) and the external inputs create the asymmetry that moves the bump along. It is difficult to visualize this asymmetry at an individual neuron level because of fluctuation, necessitating this type of population-level measure. While the mean synaptic interactions in sparsely PINned networks cause the formation of the localized bump of excitation, it is the non-trivial interaction of the fluctuations in these synaptic interactions with the external inputs that causes the bump to move across the network. Therefore, this is a novel circuit mechanism for non-autonomous sequence propagation in a network.

Additionally, we looked at other PINning-induced trends in the elements of *J$_{PINned, 8\%}$* more directly by plotting the synaptic weight values directly (Supplemental Figure 5A). The changed elements in *J$_{PINned, 8\%}$* (orange dots) were scattered away from the identity line and skewed toward more negative values. There were $pN^2$ = 40,000 of these, corresponding to the 8% plastic synaptic weights, while the other 92% of the weights remained unchanged from their initial random values (the eigenvalue spectra are shown in Supplemental Figure 5B). For all the 3 connectivity matrices of sequential networks, there was a substantial increase in the magnitudes spanned by the synaptic weights as *p* decreased (color-bars in Figures 3A and 3D), These increases in magnitude were manifested in PINned networks built with different *p* values in different ways (Figure 3F). The initial weight matrix, *J$_{Rand}$* had 0 mean (by construction), 0.005 variance (generally of order $1/N$, Experimental Procedures 1), 0 skewness and 0 kurtosis. The partially structured matrix, *J$_{PINned, 8\%}$*, on the other hand, had a negative mean of –0.1, variance of 2.2, skewness at –2, and kurtosis of 30, all of which were indicative of a probability distribution that was asymmetric about 0 and had heavy tails from a small number of strong weights. This corresponds to a network in which the large sequence-facilitating synaptic changes come from a small fraction of the weights, as suggested in experimental measurements [Song et al, 2005]. In *J$_{PINned, 8\%}$*, the ratio of the size of the largest synaptic weight to the size of the "typical" is ~20. If we assume the typical synapse corresponded to a post-synaptic potentiation (PSP) of 0.05mV, then the "large" synapses had a 1mV PSP. This is within the range in which existing experimental data support the plausibility of the network [Song et al, 2005]. For comparison purposes, the connectivity matrix for a fully structured network, *J$_{PINned, 100\%}$*, had a mean = 0, variance = 0.7, skewness = –0.02, and kurtosis = 0.2, corresponding to a network in which the synaptic changes responsible for sequences were numerous and distributed throughout the network.

Finally, we determined that synaptic connectivity matrices obtained by PINning are fairly sensitive to small amounts of structural noise, i.e., perturbations in the matrix *J$_{PINned, 8\%}$*. However, when stochastic noise (described in Experimental Procedures 2) is used during training, slightly more robust networks are obtained (Supplemental Figure 11).

# 4. Delayed Paired Association And Working Memory Can Be Implemented Through Sequences In PINned Networks

Delayed Paired Association (DPA) tasks, such as the two-alternative forced-choice (2AFC) task from [Harvey, Coen & Tank, 2012], engage working memory during each trial because the mouse must remember the identity of the cue, or cue-associated motor response, as it runs through the T-maze. DPA tasks also engage declarative memory across trials, because the mouse must remember whether each cue stimulus is associated with a left or a right turn. Therefore, in addition to being behavioral paradigms that produce sequential neural activity [Harvey, Coen & Tank, 2012], DPA tasks are useful for exploring the different neural correlates of short-term memory [Gold and Shadlen, 2007; Brunton et. al., 2013; Hanks et. al., 2015; Amit, 1995; Amit & Brunel,1995; Hansel & Mato, 2001; Hopfield & Tank, 1985; Shadlen & Newsome, 2001; Harvey, Coen & Tank, 2012]. By showing that the partially structured network we constructed by PINning could accomplish a 2AFC task, we argued that sequences can mediate an alternative form of short term memory.

During the first third of the 2AFC experiment [Harvey, Coen & Tank, 2012], the mouse received either a left or a right visual cue, and during the last third, it had two different experiences depending on whether it made a left turn or a right turn. Therefore, we modeled the cue period and the turn period of the maze by two different time-varying inputs. In the middle third, when the left- or right-specific visual cues were off (blue and red traces in Figure 5C), the mouse ran through a section of the maze that was visually identical for both types of trials for a duration equal to the delay period. In our simulation of the 2AFC task, the inputs to individual network neurons coalesced into the same time-varying waveform during the delay periods of both the left and the right trials (purple traces in Figure 5C). The correct execution of this type of task therefore depended on the network generating more than one sequence – in this case, a left sequence or a right sequence, which maintained the memory of the identity of the visual cue during a period in the task when the sensory inputs were identical.

A network with only $p$ = 16% plastic synapses generated outputs that were consistent with experimental data (Figures 5D and E, $pVar$ = 85%, $bVar$ = 40%, also compare with Figure 2c in [Harvey, Coen & Tank, 2012], here, $N$ = 569, 211 network neurons selected at random to activate in the left trial condition, schematized in blue in Figure 5A; 226 to fire in the right sequence, red in Figure 5A; and the remaining 132 to fire in the same order in both left and right sequences, non-choice-specific neurons, depicted in green in Figure 5A). This network retained the memory of cue identity by silencing the left preferring network neurons during the delay period of a right trial, and the right preferring network neurons, during a left trial, and generating sequences with the active neurons. Non-choice-specific neurons, on the other hand, were sequentially active in the same order in trials of both types, like real no-preference PPC neurons observed experimentally (outputs shown in Figure 5E, see also Supplemental Figure 7b in [Harvey, Coen & Tank, 2012]).

**5. Comparison With Fixed Point Memory Networks**

Are sequences a comparable alternative to fixed point models commonly used for storing memories? To test this idea, we compared two types of sequential-memory networks with a fixed point memory network (Figure 5F, see also Supplemental Figure 6). As before, different fractions of plastic synapses, controlled by $p$, were embedded by PINning against different target functions (Experimental Procedures 3 and 4). Here, these targets represented the values of a variable being stored and were chosen based on the dynamical mechanism by which memory is implemented – idealized for the sequential memory network (orange in Figure 5F, based on Supplemental Figure 4A), firing rates extracted from PPC data [Harvey, Coen & Tank, 2012] for the PPC-like delayed paired association network (green in Figure 5F, outputs from the PPC-like DPA network with $p$ = 16% plasticity are shown in Figure 5E), and constant valued targets for the fixed point-based memory network (blue in Figure 5F, see also Supplemental Figure 6).

To compare the task performance of the three types of memory networks under consideration here, we computed a selectivity index (Experimental Procedures 9, similar to the one used in Figure 4 in [Harvey, Coen & Tank, 2012]). We found that the network exhibiting long-duration population dynamics and memory activity through idealized sequences (orange triangles in Figure 5F) had a selectivity = 0.91, when only 10% of its synapses were modified by PINning ($p$ = 10%). In comparison, the PPC-like DPA network (whose outputs are shown in Figure 5E) needed $p$ = 16% of its synapses to be structured to match data [Harvey, Coen & Tank, 2012] and to achieve a selectivity = 0.85 (shown in the green triangles in Figure 5F). We compared the performance of fixed point memory networks relative to both these sequential memory networks, and found that both sequential memory networks performed comparably with the fixed point network in terms of their selectivity-$p$ relationship. The fixed point memory network achieved an asymptotic selectivity of 0.81 for $p$ = 23%. The magnitude of synaptic change required (Experimental Procedures 8) was also comparable between sequential and fixed point memory networks, suggesting therefore that sequences may be a viable alternative to fixed points as a mechanism for storing memories in neural circuits.

During experiments, the fraction of trials on which the mouse makes a mistake is about 15-20% (accuracy of the performance of mice at the 2AFC task was found to 83 ± 9% correct [Harvey, Coen & Tank, 2012]). We interpret errors as arising from trials in which PPC delay period activity failed to retain the identify of the cue leading to chance performance at the time that the animal makes a turn. Given a 50% probability of turning in the correct direction by chance, this implies that the cue identity is forgotten on 30-40% of trials. Adding noise to our model, we can reproduce this level of delay period forgetting with a noise amplitude of $\eta_c = 0.4 - 0.5$ (Supplemental Figure 10B). It should be noted that this noise value is in a region of noise levels where the model shows a fairly abrupt decrease in performance.

## 6. Capacity Of Sequential Memory Networks

Thus far, we have discussed one specific instantiation of delayed paired association through a sequence-based memory mechanism – the 2AFC task. Next, we extended the same basic PINning framework for constructing networks, to generate and maintain several sequential patterns of population activity. We asked whether PINned networks could accomplish memory-based tasks that required the activation of multiple, i.e., >2, non-interfering sequences. Additionally, we computed the capacity of such multi-sequential networks (denoted by $N_s$) as a function of the parameters of the networks we model, specifically, network size ($N$), PINning fraction ($p$), fraction of non-choice-specific neurons ($N_{Non-choice-specific}/N$), and temporal sparseness (fraction of sequential neurons active at any instant in the task, characterized by $N_{Active}/N$, Experimental Procedures 10).

We adjusted the PIN parameters to simulate a memory task mediated by multiple non-interfering sequences (Figure 6). Different fractions of synaptic weights in an initially random network were trained by PINning to match different target functions (schematized in Figure 6A), non-overlapping sets of Gaussian curves (similar to Supplemental Figure 4A), each of which was evenly spaced so that, collectively, they spanned a duration of 8s. The width of these waveforms, defined in terms of $N_{Active}/N$, was varied as a parameter that controlled the sparseness of the sequence (Experimental Procedures 10). Because the turn period is omitted here in the interest of clarity, the total duration of multi-sequential memory tasks modeled here is 8s-long. Similar to the DPA task modeled before (Figure 5), each network neuron received a different filtered white noise input for each cue during the cue period (0–4s), but during the delay period (4–8s), these inputs coalesced to a common cue-invariant waveform, albeit a different one for each neuron. In the most general case, we assigned $N/N_s$ neurons to each sequence that we wanted the network to produce. Here, $N_s$ is the number of memories (capacity), which also equals the number of "trial types" or the number of "cue preferences". Once again, only $p$% of the synapses in the network were plastic.

A correctly executed multi-sequential task is one in which during the delay period (here 4–8s), network neurons fire in a sequence only on trials of the same type as their cue preference, and are silent during other trials. For example, the set of neurons selective for Cue 1 fires in a sequence only during Trial Type 1 and not during Trial Types numbering 2 – $N_s$, neurons selective for Cue 2 activate only during Trial Type 2, and so on (schematized in Figure 6A). A network of 500 neurons with $p$ = 25% plastic synapses performed such a task easily, generating $N_s$ = 5 sequences with delay period memory of the appropriate cue identity with temporal sparseness of $N_{Active}/N$ = 3% of the total number of network neurons (Figure 6B).

Temporal sparseness of the sequences was found to be a crucial factor that determined how well a network performed a multi-sequential memory task (Experimental Procedures 10). When sequences were forced to be sparser than a certain minimum (by using narrower target waveforms, for this particular 8s-long task, this occurred when $N_{Active}/N$ < 1.6%) the network failed. Although there was sequential activation, the memory of the identity of the 5 separately memorable cues was not maintained across the delay period. Surprisingly, we found that this failure could be rescued by adding a small number of non-choice-specific neurons ($N_{Non\text{-}choice\text{-}specific}/N$ = 4%) that fired in the same temporal order in all 5 trial types (Figure 6C). In other words, the remaining $N - N_{Non\text{-}choice\text{-}specific}$ network neurons (< $N$) now successfully executed the memory-based multi-sequential task using 5 sparse choice-specific sequences and 1 non-choice-specific sequence. Thus, the capacity of the network for producing multiple temporally sparse sequences increased when non-choice-specific neurons were present to stabilize them, without requiring a concomitant increase in the amount of synaptic modification. Therefore, we predict that non-choice-specific neurons, also seen in the PPC [Harvey, Coen & Tank, 2012], may function as a "conveyor belt" of working memory, providing recurrent synaptic current to sequences that may be too sparse to sustain themselves otherwise in a non-overlapping scheme.

Finally, we focused on two of these aspects – memory capacity and noise tolerance. The presence of non-choice-specific neurons increased the capacity of networks to store memories if they were implemented through sequences (Figure 6C). Up to a constant factor, the memory capacity, the maximum of the fraction $N_s/N$, scaled in proportion to the fraction of plastic synapses, $p$ and the network size, $N$, and inversely with the sparseness, $N_{Active}/N$ (Figure 6D). The slope of this capacity-to-network-size relationship increased when non-choice-specific neurons were included because they enabled the network to carry sparser sequences. Furthermore, we found that networks with a bigger fraction of plastic synaptic connections and networks containing non-choice-specific neurons were more stable against stochastic perturbations (Figure 6E). Once the amplitude of added stochastic noise exceeded the maximum tolerance of a particular network (denoted by $\eta_c$, Experimental Procedures 2), however, non-choice-specific neurons were no longer effective at repairing the memory capacity (not shown); non-choice-specific neurons could not rescue inadequately PINned multi-sequential schemes either (small $p$, not shown).

**DISCUSSION**

In this paper, we used and extended earlier work based on liquid-state [Maass, Natschlager & Markram, 2002] and echo-state machines [Jaeger, 2003], which has shown that a basic balanced-state network with random recurrent connections can act as a general purpose dynamical reservoir, the modes of which can be harnessed for performing different tasks by means of feedback loops. These models typically compute the output of a network through weighted readout vectors [Buonomano & Merzenich, 1995; Maass, Natschlager & Markram, 2002; Jaeger, 2003; Jaeger & Haass, 2004] and feed the output back into the network as an additional current, leaving the synaptic weights within the dynamics-producing network unchanged in their randomly initialized configuration. The result was that networks generating chaotic spontaneous activity prior to learning [Sompolinsky, Crisanti & Sommers, 1988], produced a variety of regular non-chaotic outputs that match the imposed target functions after learning [Sussillo & Abbott, 2009]. The key to making these ideas work is that the feedback carrying the error during the learning process forces the network into a state in which it produces less variable responses [Molgedey, Schuchhardt & Schuster, 1992; Bertschinger & Natschlager, 2004; Rajan, Abbott & Sompolinsky, 2010; Rajan, Abbott & Sompolinsky, 2011]. A compelling example, FORCE learning [Sussillo & Abbott, 2009], has been implemented as part of several modeling studies [Laje & Buonomano, 2013; Sussillo, 2014], and references therein), but it had a few limitations. Specifically, the algorithm in [Sussillo & Abbott, 2009] included a feedback network separate from the network generating the actual trajectories, which contained aplastic connections in a black-box, and learning was restricted only to a set of readout weights. The approach we developed here, called Partial In-Network Training or PINning, avoids these issues, while propagating neural sequences resembling data (Sections 1 and 2). During PINning, only a fraction of the recurrent synapses in an initially random network are modified in a task-specific manner and by a biologically reasonable amount, leaving the majority of the network heterogeneously wired. Furthermore, the fraction of plastic synaptic connections in these networks is a tunable parameter that controls the contribution of the structured portion of the network relative to the random part, allowing these models to interpolate smoothly between highly structured and completely random architectures until we find the point that best matches the relevant experimental data (a similar point is made in Barak et al, 2013). PINning is not the most general way to restrict learning to a limited number of synapses, but it allows us to do so without losing the efficiency of the learning algorithm.

To illustrate the applicability of this framework for constructing partially structured networks, we used experimental data recorded from the posterior parietal cortex (PPC) of mice performing a working memory-based decision making task in a virtual reality environment [Harvey, Coen & Tank, 2012]. It is unlikely that a high-order association cortex such as the PPC evolved specialized circuitry solely for sequence generation, especially considering that the PPC appears to retain its ability to mediate other complex temporal tasks, from working memory and decision making to evidence accumulation and navigation [Shadlen & Newsome, 2001; Gold & Shadlen, 2007; Freedman & Assad, 2011; Snyder, Batista & Anderson, 1997; Anderson & Cui, 2009; Bisley & Goldberg, 2003; McNaughton, 1994; Nitz, 2006; Whitlock et al, 2008; Carlton & Taube, 2009; Hanks et al, 2015]. We also computed a measure of stereotypy from PPC data (*bVar*, Figure 1C–G) and found it to be much lower than if the PPC were to generate sequences based on highly specialized intrinsic connectivity (for example, with chain-like or ring-like connections) or if it merely read out sequential activity from a highly structured upstream region. We therefore started with random recurrent networks, imposed a small amount of structure in their connectivity through PINning and duplicated many features of the experimental data [Harvey, Coen & Tank, 2012], primarily among them, choice-specific neural sequences and retention of the memory of cue identity during the delay period.

We analyzed the the structural features in the synaptic connectivity matrix of sparsely PINned networks, concluding that the probability distribution of the synaptic strengths is heavy tailed due to the presence of a small percentage of strong interaction terms (Figure 3F). There is experimental evidence that there might be a small fraction of very strong synapses embedded within a milieu of a large number of relatively weak synapses in the cortex [Song et al, 2005], most recently, from the primary visual cortex [Cossell & Mrsc-Flogel, 2015]. Synaptic distributions measured in slices have been shown to have long tails [Song & Nelson, 2005], and the experimental result in [Cossell & Mrsc-Flogel, 2015] has demonstrated that rather than occurring at random, these strong synapses significantly contribute to network tuning by preferentially interconnecting neurons with similar orientation preferences. These strong synapses may be the plastic synapses that are induced by PINning in our scheme. The model exhibits some structural noise sensitivity (Supplemental Figure 11), and this is not completely removed by training in the presence of noise. It is possible that dynamic mechanisms of ongoing plasticity could enhance stability to structural fluctuations.

In this paper, the circuit mechanisms underlying both the *formation* of a localized bump of excitation in the connectivity, and the manner in which the bump of excitation *propagates* across the network, were elucidated. For the first, we quantified the influence of network neurons that are away from the ones active in the sequence (Figure 3). We found that the mechanism for bump formation in the networks constructed by sparse PINning is consistent with "center surround"-like features in the mean synaptic interactions (Figure 3B, see also Figure 3C), as expected [Yishai, Bar-Or & Sompolinsky, 1995; Zhang, 1996]. For the second, we analyzed the collective impact of recurrent synaptic currents and external inputs onto an example neuron active at one specific time point in the sequence (Figure 4). From these results, we concluded that the circuit mechanism for the propagation of the sequence is non-autonomous, relying on a complex interplay between the connections in these networks and the external inputs. The mechanism for bump propagation is therefore distinct from the standard moving bump/ring attractor model [Yishai, Bar-Or & Sompolinsky, 1995; Zhang, 1996], but has similarities to the models developed in [Fiete et al, 2010; Hopfield, 2015] (explored in Supplemental Figure 9). In particular, the model in [Fiete et al, 2010] successfully generates highly stereotyped and noise-free sequences, similar to those observed in area HVC experimentally. This model is initialized as a recurrently connected network similar to ours and subsequently uses spike-timing dependent plasticity, heterosynaptic competition and correlated external inputs to learn sequential activity. A critical difference between [Fiete et al, 2010] and the approach presented here lies in the underlying network connectivity responsible for sequences – their learning rule results in synaptic chain-like connectivity.

We suggest an alternative hypothesis that sequences might be a more general and effective dynamical form of working memory (Sections 4 and 5), making the prediction that sequences may be observed in many experiments involving diverse tasks that require working memory (also suggested in the Discussion section of [Harvey, Coen & Tank, 2012]). This contrasts with previous models of working memory that relied on fixed point attractors to retain information and exhibited sustained activity [Amit, 1992; Amit, 1995; Amit & Brunel,1995; Hansel & Mato, 2001; Hopfield & Tank, 1985]. Finally, we computed the capacity of sequential memory networks for storing more than two memories (Section 6) by extending the same sparse PINning approach developed for matching the data from a 2AFC experiment [Harvey, Coen & Tank, 2012] in Section 4. Experiments on freely behaving animals performing more complex tasks, e.g., with more than two contingencies, will likely show that more than two memories can be stored through sequences. Furthermore, ongoing experiments in several laboratories are testing the effect of switching the contextual meaning of multiple cues and cue-combinations, including olfactory, visual, and multi-sensory cues, in basic memory-guided decision-making tasks and in complex variants of such tasks. Analyzing population data from such studies will soon allow experimental tests of the multi sequential network models constructed by PINning.

The capacity of the sequential memory networks constructed by PINning can be thought of the "computational bandwidth" of a general-purpose neural circuit to perform different timing-based computations. Here, we showed that a small amount of structured connections embedded in a much larger skeleton of disordered connections is sufficient for sequential timing signals of the order of 10s. We can also ask, what else the rest of the network can do. A network could be channeled by sparse training to first perform two (or more) different temporally structured tasks, for instance, accumulation of evidence ([Brunton, Botvinick & Brody, 2013; Hanks et al, 2015], and references therein) and delayed paired association ([Harvey, Coen &

Tank, 2012], and references therein), and then switch between these tasks in a context-dependent manner. In the future, using the same analyses developed in this paper on such "multi-purpose" networks could make biological predictions about circuit-level mechanisms operating in areas such as the PPC that are implicated in both these, as well as in other tasks [Shadlen & Newsome, 2001; Gold & Shadlen, 2007; Freedman & Assad, 2011; Snyder, Batista & Anderson, 1997; Anderson & Cui, 2009; Bisley & Goldberg, 2003; McNaughton, 1994; Nitz, 2006; Whitlock et al, 2008; Carlton & Taube, 2009; Hanks et al, 2015].

The term "pre-wired" is used in this paper to mean a scheme in which a principled mechanism for executing a certain task is first assumed, and then incorporated into the network circuitry, for example, a moving bump architecture [Yishai, Bar-Or & Sompolinsky, 1995; Zhang, 1996], or a synfire chain [Hertz & Prugel-Bennett,1996; Levy et al, 2001; Hermann, Hertz & Prugel-Bennet, 1995; Fiete et al, 2010]. In contrast, the models built and described in this paper are constructed without bias or assumptions. If a moving bump architecture [Yishai, Bar-Or & Sompolinsky, 1995; Zhang, 1996] had been assumed at the beginning and the network pre-wired accordingly, we would of course have uncovered it through the analysis of the synaptic connectivity matrix (similar to Section 3, Figures 3 and 4 and Supplemental Figure 9, see also Experimental Procedures 11). However, by starting with an initially random configuration and learning a small amount of structure, we found an alternative mechanism for input-dependent sequence propagation (Figure 4). We would not have encountered this mechanism for the non-autonomous movement of the bump by pre-wiring a different mechanism into the connectivity of the model network. While the models constructed here are indeed trained to perform the task, the fact that they are unbiased means that the opportunity was present to uncover mechanisms that were not thought of *a priori*.

# EXPERIMENTAL PROCEDURES

## 13.1. Network elements

We consider a network of *N* fully interconnected neurons described by a standard firing rate model. Each model neuron is characterized by an activation variable, $x_i$ for *i = 1, 2, ... N*, where, *N* = 437 for the PPC-like sequence in Figures 1D and 2A, *N* = 500 for the single idealized sequence in Supplemental Figure 4 and the multi-sequential memory task in Figure 6, and *N* = 569 for the 2AFC task in Figure 5 (we generally build networks of the same size as the experimental dataset we are trying to model, however the results obtained remain applicable to larger networks, see for example, Supplemental Figure 7), and a nonlinear response function, $\phi(x) = \dfrac{1}{1+e^{[-(x-\theta)]}}$. This function ensures that the firing rates, $r_i = \phi(x_i)$, go from a minimum of 0 to a maximum at 1. Adjusting $\theta$ allows us to set the firing rate at rest, *x* = 0, to some convenient and biologically realistic background firing rate, while retaining a maximum gradient at $x = \theta$. We use $\theta = 0$ (but see also Supplemental Figure 8).

We introduce a recurrent synaptic weight matrix **J** with element $J_{ij}$ representing the strength of the connection from presynaptic neuron *j* to postsynaptic neuron *i* (schematic in Figure 1A) The individual synaptic weights are initially chosen independently and randomly from a Gaussian distribution with mean and variance given by $\langle J_{ij} \rangle_J = 0$ and $\langle J_{ij} \rangle_J^2 = g^2 / N$, and are either held fixed or modifiable, depending on the fraction of plastic synapses *p* that can change by applying a learning algorithm (Experimental Procedures 4).

The activation variable for each network neuron $x_i$ is determined by,

$$\tau \frac{dx_i}{dt} = -x_i + \sum_j^N J_{ij} \phi(x_j) + h_i .$$

In the above equation, $\tau$ = 10ms is the time constant of each unit in the network and the control parameter *g* determines whether (*g* > 1) or not (*g* < 1) the network produces spontaneous activity with non-trivial dynamics [Sompolinsky, Crisanti & Sommers, 1988; Rajan, Abbott & Sompolinsky, 2010; Rajan, Abbott & Sompolinsky, 2011]. We use *g* values between 1.2 and 1.5 for the networks in this paper, so that the randomly initialized network generates chaotic spontaneous activity prior to activity-dependent modification (gray trace in the schematic in Figure 2B), but produces a variety of regular non-chaotic outputs that match the imposed target functions afterward (red trace in the schematic in Figure 2B, see also results in Figures 2, 5 and 6, see also Experimental Procedures 4 later). The network equations are integrated using Euler method with an integration time step, *dt* = 1ms. $h_i$ is the external input to the unit *i*.

### 13.2. Design of External Inputs

During the course of the real two-alternative forced-choice (2AFC) experiment [Harvey, Coen & Tank, 2012], as the mouse runs through the virtual environment, the different patterns projected onto the walls of the maze (colored dots, stripes, pillars, hatches, etc.) translate into time-dependent visual inputs arriving at the PPC. Therefore, to represent sensory (visual and proprioceptive) stimuli innervating the PPC neurons, the external inputs to the neurons in the network, denoted by $h(t)$, are made from filtered and spatially delocalized white noise that is frozen (repeated from trial to trial), using the equation,

$\tau_{WN}\frac{dh}{dt} = -h(t) + h_0 \eta(t)$, where $\eta$ is a random variable drawn from a Gaussian distribution with 0 mean and unit variance, and the parameters $h_0$ and $\tau_{WN}$ control the scale of these inputs and their correlation time, respectively. We use $h_0 = 1$ and $\tau_{WN} = 1$ s. There are as many different inputs as there are model neurons in the network, with individual model neurons receiving the same input on every simulated trial. A few example inputs are shown in the right panel of Figure 1A.

In addition to the frozen noise *h*, which acts as external inputs to these networks, described above, we also test the resilience of the memory networks we built (Figure 6E) to injected stochastic noise. This stochastic injected noise varies randomly (i.e., is a Gaussian random variable between 0 and 1, drawn from a zero mean and unit variance distribution) and independently at every time step. The diffusion constant of the white noise is given by $A_\eta^2/2\tau$, where the amplitude is $A_\eta^2$ and $\tau$ is the time constant of the network units (we use 10ms, as detailed in Experimental Procedures 1). We define "Resilience" or "Noise Tolerance" as the critical amplitude of this stochastic noise, denoted by $\eta_c$, at which the delay period memory fails and the Selectivity Index of the memory network drops to 0 (Figure 6E, see also Supplemental Figure 10).

### 13.3. Extracting Target Functions From Calcium Imaging Data

To derive the target functions for our activity-dependent synaptic modification scheme termed Partial In-Network Training or PINning, we convert the calcium fluorescence traces from PPC recordings [Harvey, Coen & Tank, 2012] into firing rates using two complementary methods. We find that for this dataset, the firing rates estimated by the two methods agree quite well (Supplemental Figure 1).

The first method is based on the assumption that the calcium impulse response function, which is a difference of exponentials ($K \propto e^{-t/384} - e^{-t/52}$, with a rise time of 52ms and a decay time of 384ms [Tian et al, 2009; Harvey, Coen & Tank, 2012]), is approximated by an alpha function of the form $K \propto te^{-t/\tau_{Ca}}$, where there is only a single (approximate) time constant for the filter, $\tau_{Ca}$ = 200ms. According to this assumption, the scaled firing rate *s* and calcium concentration, *[Ca²⁺]* are related by,

$$\tau_{Ca}\frac{dCa_i}{dt} = -Ca_i(t) + x_i(t) \text{ and } \tau\frac{dx_i}{dt} = -x_i(t) + s_i(t),$$

where, x(t) is an auxiliary variable. The inverse of the above model is obtained by taking a derivative of the calcium data, writing,

$$x_i(t) = Ca_i(t) + \tau_{Ca} \frac{dCa_i}{dt} \text{ and } s_i(t) = x_i(t) + \tau \frac{dx_i}{dt}.$$

Once we have *s(t)*, we rectify it and choose a smoothing time constant $\tau_R$ for the firing rate we need to compute. Finally, integrating the equation, $\tau_R \frac{dR_i}{dt} = -R_i(t) + s_i(t)$, and normalizing by the maximum gives us an estimate for the firing rates extracted from the calcium data, denoted by *R* (Supplemental Figure 1).

The second method is a fast Bayesian deconvolution algorithm [Pnevmatikakis et al, 2014; Vogelstein et al, 2010, available online at https://github.com/epnev/continuous_time_ca_sampler] that infers spike trains from calcium fluorescence data. The inputs to this algorithm are the rise time (52ms) and the decay time (384ms) of the calcium impulse response function [Tian et al, 2009; Harvey, Coen & Tank, 2012] and a noise parameter [Pnevmatikakis et al, 2014; Vogelstein et al, 2010]. Typically, if the frame rate for acquiring the calcium images is low enough (the data in [Harvey, Coen & Tank, 2012] are imaged at 64ms per frame), the outputs from this algorithm can be interpreted as a normalized firing rate. To verify the accuracy of the firing rate outputs obtained from trial-averaged calcium data (for example, from Figure 2c in [Harvey, Coen & Tank, 2012]), we smoothed the spike trains we got from the above method for each trial separately through a Gaussian of the form $\sum_i e^{\frac{-(t-t_i)}{2\tau_R^2}}$, normalized by $\sqrt{2\pi} \times \tau_R$, and then averaged over single trials to get trial-averaged firing rates (this smoothing and renormalization procedure has also been recommended for faster imaging times [Pnevmatikakis et al, 2014; Vogelstein et al, 2010]).

Once the values of $\tau_R$ and $\tau_{Ca}$ are determined that make the results obtained by both deconvolution methods consistent (we used $\tau_R$ = 100ms and $\tau_{Ca}$ = 384ms), we used the firing rates extracted as target functions for PINning through the transform, $f_i(t) = \ln\left[\frac{R_i(t)}{1 - R_i(t)}\right]$. The above expression is obtained by solving the activation function relating input current to firing rate of model neurons, $R_i(t) = \frac{1}{1 + e^{-f_i(t)}}$, since the goal of PINning is to match the input to neuron *i*, say, denoted by $z_i(t)$ to its target function, denoted by $f_i(t)$. Finally, to verify our estimates, we re-convolved (Supplemental Figure 1) the output firing rates from the network neurons with a difference of exponentials using a rise time of 52ms and a decay time of 384ms.

### 13.4. Synaptic Modification Rule For PINning

During PINning, the inputs of individual network neurons are compared directly with the target functions to compute a set of error functions, i.e., $e_i(t) = z_i(t) - f_i(t)$, for *i = 1, 2, ... N*. Individual neuron inputs are expressed as $z_i(t) = \sum_j J_{ij} r_j(t)$, where $r_j(t)$ is the firing rate of the *j*th or the presynaptic neuron.

During learning, the subset of plastic internal weights in the connectivity matrix **J** of the random recurrent network, denoted by the fraction *p*, undergo modification at a rate proportional to the error term, the presynaptic firing rate of each neuron, $r_j$ and a *pN x pN* matrix, **P** (with elements $P_{ij}$) that keeps track of the rate fluctuations across the network at every time step. Here, *p* is the fraction of neurons whose outgoing synaptic weights are plastic; since this is a fully connected network, this is also the fraction of plastic synapses in the network. Mathematically, $P_{ij} = <r_i r_j>^{-1}$, the inverse cross-correlation matrix of the firing rates of the network neurons (*P<sub>ij</sub>* is computed for all *i* but is restricted to *j = 1, 2, 3, …,pN*). The basic algorithm is schematized in Figure 2B. At time *t*, for *i = 1, 2, ... N* neurons, the learning rule is simply that the elements of the matrix **J** are moved from their values at a time step $\Delta t$ earlier through $J_{ij}(t) = J_{ij}(t-1) + \Delta J_{ij}(t)$. Here, the synaptic update term, according to the RLS/FORCE procedure [Haykins, 2002; Sussillo & Abbott, 2009] (since other methods for training recurrent networks, such as backpropagation would be too laborious for our purposes) follows,

$\Delta J_{ij}(t) = c[z_i(t) - f_i(t)] \sum_k P_{jk}(t) r_k(t)$, where the above update term is restricted to the *p*% of plastic synapses in the network, which are indexed by *j* and *k* in the above expression. While c can be thought of as an effective learning rate, it is given by the formula, $c = \dfrac{1}{1 + r'(t)\mathbf{P}(t)r(t)}$. The only free parameter in the learning rule is **P(0)** (but the value to which it is set is not critical [Sussillo & Abbott, 2009]). When there are multiple sequences (such as in Figures 5 and 6), we choose *pN* synapses that are plastic and we use those same synapses for all the sequences.

The matrix **P** is generally not explicitly calculated but rather updated according to the rule,

$\mathbf{P}(t) = \mathbf{P}(t-1) - \dfrac{\mathbf{P}(t-1)r(t)r'(t)\mathbf{P}(t-1)}{1 + r'(t)\mathbf{P}(t-1)r(t)}$ in matrix notation, which includes a regularizer [Haykins, 2002]. In our scheme, all indices in the above expression are restricted to the neurons with plastic synapses in the network. The algorithm requires the matrix **P** to be initialized to the identity matrix times a factor that controls the overall learning rate, i.e., $P(0) = \alpha \times I$, and in practice, values from 1 to 10 times the overall amplitude of the external inputs (denoted by *h<sub>0</sub>* in Experimental Procedures 2) driving the network are effective (other values are explored in [Sussillo & Abbott, 2009]).

For numerically simulating the PINned networks whose sequential outputs are shown in Figures 2C, 5 and 6, the integration time step used is *dt* = 1ms (as described in Experimental Procedures 1 and 2, we use Euler method for integration). The learning occurs at every time step for the *p*% of pre-synaptic neurons with plastic outgoing synaptic weights. Starting from a random initial state (Experimental Procedures 1), we first run the program for 500 learning steps, which include both the network dynamics and the PINning algorithm, and then an additional 50 steps with only the network dynamics after the learning has been terminated (convergence metrics below, see also Supplemental Figure 2). A "step" is defined as one run of the program for the duration of the relevant trial, denoted by *T*. Each step is equivalent to *T* = 10500 time points (10.5s) in Figures 2C and 5E, and 8000 time points (8s) in Figure 6B.

On a standard laptop computer, the first 500 such steps for a 500-unit rate-based network producing a 10s-long sequence (such as in Supplemental Figure 4) take approximately 8 minutes to complete in realtime and the following 50, about 30 seconds in realtime (about 1s/step for the 10s-long single sequence example, scaling linearly with network size *N*, total duration or length of the trial *T*, and number of sequences produced.)

The convergence of the PINning algorithm was assayed as follows: (a) By directly comparing the outputs with the data (as in the case of Figures 2A–C and 5D–E) or the set of target functions used (Supplemental Figure 4) and (b) By calculating and following the $\chi^2$-squared error between the network rates and the targets, both during PINning (inset of Supplemental Figure 2) and at the end of the simulation (Supplemental Figure 2). The performance of the PINning algorithm was assayed by computing the percent variance of the data or the targets captured by the sequential outputs of different networks (*pVar*, see Experimental Procedures 7, see also Figure 1H) or by computing the Selectivity Index of the memory network (Selectivity, see Experimental Procedures 9, see also Figure 5F).

### 13.5. Dimensionality Of Network Activity ($Q_{eff}$)

We use state space analysis based on PCA (see for example, [Rajan, Abbott & Sompolinsky, 2011; Sussillo, 2014] and references therein) to describe the instantaneous network state by diagonalizing the equal-time cross-correlation matrix of network firing rates given by, $Q_{ij} = \langle (r_i(t) - \langle r_i \rangle)(r_j(t) - \langle r_j \rangle) \rangle$, where <> denotes a time average. The eigenvalues of this matrix expressed as a fraction of their sum indicate the distribution of variances across different orthogonal directions in the activity trajectory. We define the effective dimensionality of the activity, $Q_{eff}$, as the number of principal components that capture 95% of the variance in the dynamics (Figure 2F).

### 13.6. Stereotypy Of Sequence (*bVar*), %

*bVar* quantifies the variance of the data or the network output that is explained by the translation along the sequence of an activity profile with an invariant shape. For example, for Figure 1C–G, we extracted an aggregate waveform, denoted by $R_{ave}$ (red trace in Figures 1D and Supplemental Figure 4A), by averaging the $t_{COM}$-realigned firing rates extracted from trial-averaged PPC data collected during a 2AFC task [Harvey, Coen & Tank, 2012]. Undoing the $t_{COM}$ shift, we can write this function for the aggregate or "typical" bump-like waveform as $R_{ave}(t-t_i)$. The amount of variability in the data that is explained by the moving bump, $R_{ave}(t-t_i)$ is given by a measure we call *bVar*.

$$bVar = \left[1 - \frac{\langle R_i(t) - R_{ave}(t-t_i)\rangle^2}{\langle R_i(t) - \bar{R}(t)\rangle^2}\right], \text{ where } \langle\rangle = \sum_i^N \sum_t^T.$$ In the above expressions, *R*'s denote the firing rates extracted from calcium data (Experimental Procedures 3); $R_i(t)$ is the firing rate of the $i^{th}$ PPC neuron at time *t* and $\bar{R}(t)$ is the average over neurons. The total duration, *T* is 10.5s in Figures 1, 2 and 5, and 8s in Figure 6.

In some ways, *bVar* is similar to the ridge-to-background ratio computed during the analysis of experimental data for measuring the level of background activity (see for example, Supplemental Figure 14 in [Harvey, Coen & Tank, 2012]); however, *bVar* additionally quantifies the stereotypy of the shape of the transient produced by individual neurons.

### 13.7. Percent Variance Of Data Explained By Model (*pVar*), %

We quantify the match between the experimental data or the set of target functions, and the outputs of the model by the amount of variance of the data that is captured by the model,

$$pVar = \left[1 - \frac{\langle D_i(t) - r_i(t)\rangle^2}{\langle D_i(t) - \bar{D}(t)\rangle^2}\right],$$ which is one minus the ratio of the Frobenius norm of the difference between the data and the outputs of the network, and the variance of the data. The data referred to here, denoted by *D*, is trial-averaged data, such as from Figures 1D, 2A and 5D.

### 13.8. Magnitude Of Synaptic Change

In Figure 2E and in Figure 5G, we compute the magnitude of the synaptic change required to implement a single PPC-like sequence, an idealized sequence and three memory tasks, respectively. In combination with the fraction of plastic synapses in the PINned network, *p*, this metric characterizes the amount of structure that needs to be imposed in an initially random network to produce the desired temporally structured dynamics. This is calculated as, normalized mean synaptic change =

$$\frac{\sum_{ij} |J_{ij}^{PINned, p\%} - J_{ij}^{Rand}|}{\sum_{ij} |J_{ij}^{Rand}|},$$ where, following the same general notation as in the main text, $J_{PINned, p\%}$

denotes the connectivity matrix of the PINned network constructed with *p*% plastic synapses and $J_{Rand}$ denotes the initial random connectivity matrix (*p* = 0).

### 13.9. Selectivity Index For Memory Task

In Figures 5F, a Selectivity Index is computed (similar to Figure 4 in [Harvey, Coen & Tank, 2012]) to assess the performance of different PINned networks at maintaining cue-specific memories during the delay period of delayed paired association tasks. This metric is based on the ratio of the difference and the sum of the mean activities of preferred neurons at the end of the delay period during preferred trials, and the mean activities of preferred neurons during opposite trials. We compute Selectivity Index as,

$$\frac{1}{2}\left[\frac{<r>_{r.t}^{r.n} - <r>_{l.t}^{r.n}}{<r>_{r.t}^{r.n} + <r>_{l.t}^{r.n}} + \frac{<r>_{l.t}^{l.n} - <r>_{r.t}^{l.n}}{<r>_{l.t}^{l.n} + <r>_{r.t}^{l.n}}\right],$$ where the notation is as follows:

$$<r>_{r.t}^{r.n} = \frac{1}{N_{\text{right pref neurons}}} \sum_{i}^{\text{right pref neurons}} r_i \text{ and } <r>_{r.t}^{l.n} = \frac{1}{N_{\text{left pref neurons}}} \sum_{i}^{\text{left pref neurons}} r_i \text{ are the average firing}$$

rates of right-preferring and left-preferring model neurons on right trials;

$$<r>_{l.t}^{r.n} = \frac{1}{N_{\text{right pref neurons}}} \sum_{i}^{\text{right pref neurons}} r_i \text{ and } <r>_{l.t}^{l.n} = \frac{1}{N_{\text{left pref neurons}}} \sum_{i}^{\text{left pref neurons}} r_i \text{ are the average firing}$$

rates of right-preferring and left-preferring model neurons on left trials of the simulated task. The end of the delay period is at approximately 10s for the network in Figure 5 (after [Harvey, Coen & Tank, 2012]) and at ~7s time point for the network in Figure 6.

### 13.10. Temporal Sparseness Of Sequences ($N_{Active}/N$)

The temporal sparseness of a sequence is defined as the fraction of neurons active at any instant during the sequence, the fraction, $N_{Active}/N$. To compute this, first, the normalized firing rate from each model neuron in the network or from data [Harvey, Coen & Tank, 2012], denoted by $R_i(t)$, is realigned by the center-of-mass, $i_{COM}(t)$, given by, $i_{COM}(t) = \sum_i i * R(i,t) / \sum_i R(i,t)$, where $i$ is the neuron number and $t$ is time. The realigned rates are then averaged over time to obtain $<R>_{time}$, i.e., $\langle R(i) \rangle_{time} = \langle R(i - i_{COM}(t), t) \rangle_{time}$ after using a circular shift rule to undo the $i_{COM}$-shift. The standard deviation of the best Gaussian that fits this curve $<R>_{time}$ is the number, $N_{Active}$, and the ratio of $N_{Active}$ to the network size, $N$, yields the temporal sparseness of the sequence, $N_{Active}/N$. For the data in Figure 2A, for example, $N_{Active}/N = 3\%$ (i.e., 16 neurons out of a total of 437). Practically speaking, decreasing this fraction makes the sequence narrower and at a critical value of sparseness ($N_{Active}/N = 1.6\%$ in Figure 6), there is not enough current in the network to propagate the sequence. However, up to a point, making the sequences sparser increases the capacity of a network for carrying multiple non-interfering sequences (i.e., sequences with delay period memory of cue identity), without demanding an increase in either the fraction of plastic synapses, $p$, or in the overall magnitude of synaptic change.

**13.11. Analyzing the Structure of PINned Synaptic Connectivity Matrices**

This is pertinent to Section 3 (Figures 3 and 4), in which we quantify how the synaptic strength varies with the "distance" between pairs of network neurons in connectivity space, $i - j$, in PINned sequential networks. We first compute the means and the standard deviations of the principal diagonals, i.e., $\sum_{i=j}^{N} \frac{J_{ij}}{N} \rightarrow \sum_{i=1}^{N} \frac{J_{ii}}{N}$, in the 3 connectivity matrices under consideration here – $J_{PINned, 8\%}$, $J_{Rand}$, and just for comparison purposes, $J_{PINned, 100\%}$. Then, we compute the means and the standard deviations of successive off-diagonal "stripes" moving away from the principal diagonal, i.e., $\sum_{i-j=a} \frac{J_{ij}}{N} \rightarrow \sum_{i=a+1}^{N} \frac{J_{i(i-a)}}{N}$, for the same three matrices. These are plotted in Figures 3B for the sparsely PINned matrix, $J_{PINned, 8\%}$, relative to the randomly initialized matrix, $J_{Rand}$, and in Figure 3E, for the fully PINned matrix constructed for comparison purposes, $J_{PINned, 100\%}$. The same analysis is also used to compare different partially structured matrices in Supplemental Figure 9.

**FIGURES AND CAPTIONS**

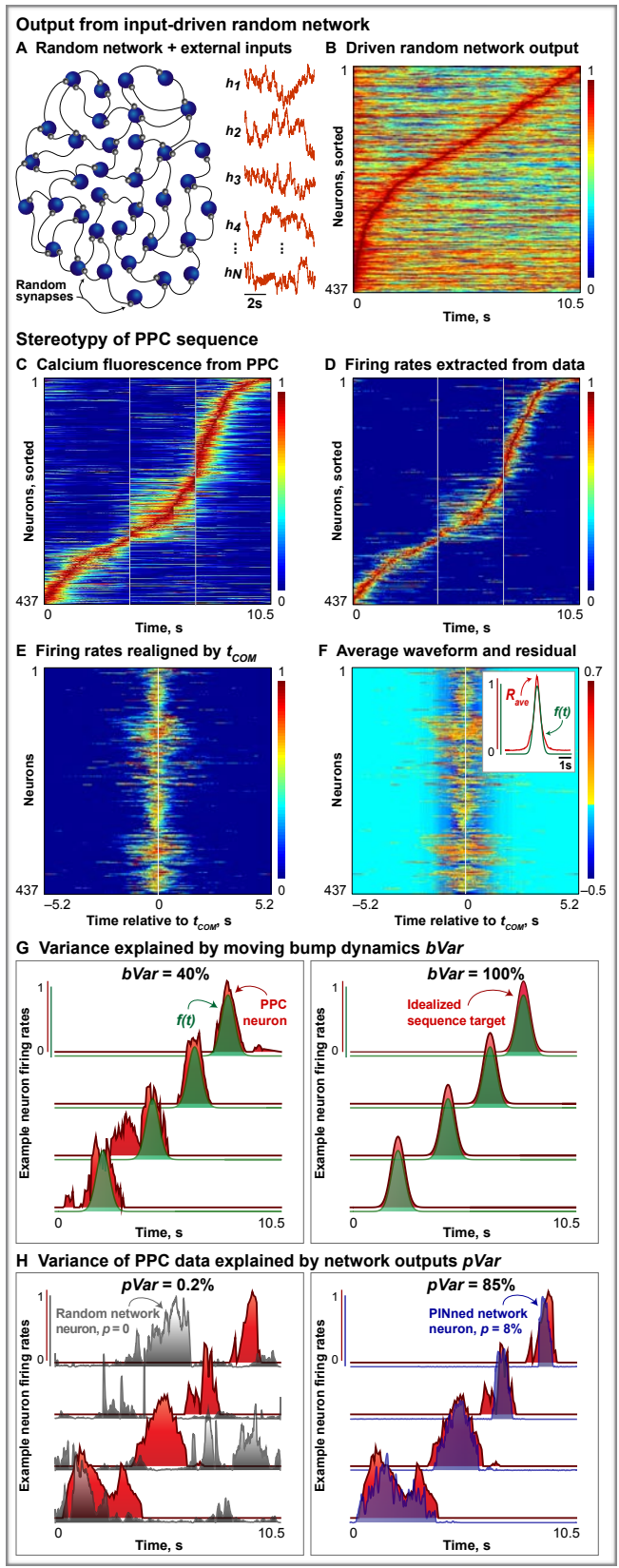

## Figure 1: Stereotypy Of PPC Sequences And Random Network Output

A. Schematic of a randomly connected model network of firing rate neurons (depicted in blue, Experimental Procedures 1) operating in a spontaneously active regime and a few examples of the temporally irregular external inputs to the network (orange traces, Experimental Procedures 2) are shown here. Random synapses are depicted in gray. Note that the networks we build contain as many rate-based model neurons as the size of the PPC dataset [Harvey, Coen & Tank 2012] under consideration, and are typically all-to-all connected, but only a fraction of these model neurons and their interconnections are depicted in these schematics.

B. When the individual firing rates of the random untrained network driven by external inputs in (A) are normalized by the maximum per neuron and sorted by their $t_{COM}$, the activity across the unstructured network appears time ordered. However, this sequence contains large amounts of extra-sequential or background activity, compared to the PPC data (*bVar* = 12%, *pVar* = 0.2%).

C. Calcium fluorescence (i.e., normalized $\Delta F/F$) data collected from 437 PPC neurons during an 10.5 second-long 2AFC experiment, corresponding to both left and right correct-choice outcomes, from 437 trial averaged neurons and pooled across 6 mice.

D. Normalized firing rates extracted from (C) using deconvolution methods (Experimental Procedures 3, see also Supplemental Figure 1) are shown here.

E. The firing rates from the 437 neurons shown in (D) are realigned by their time of center-of-mass (abbreviated $t_{COM}$) and plotted here.

F. **Inset:** A "typical" waveform ($R_{ave}$ in red) is obtained by averaging the realigned rates from (C) over neurons, and a Gaussian curve with mean = 0 and variance = 0.3 ($f(t)$, green trace), that best fits the neuron-averaged waveform, are both plotted here. **Main panel:** Residual activity level that is not explained by translations of the best fit to $R_{ave}$, $f(t)$, is shown here.

G. The variance in the population activity that is explained by translations of the best fit to $R_{ave}$, $f(t)$, is a measure of the stereotypy in the activity (abbreviated $bVar$ in the main text, Experimental Procedures 6). Left panel shows the normalized firing rates from 4 example PPC neurons (red) from (D), and the curves $f(t)$ for each example neuron (green). $bVar$ = 40% for these data. Right panel shows the normalized firing rates from 4 model neurons (red) from a network generating an idealized sequence (such as in Supplemental Figure 4A) and the corresponding curves $f(t)$ for each (green). $bVar$ = 100% for this idealized sequential network.

H. The variance of the PPC data that is explained by the outputs of the different networks constructed by PINning is given by $pVar$ (Experimental Procedures 7) and illustrated with 4 example neurons here. Left panel shows the normalized rates from 4 example PPC neurons (red) picked from (D) and 4 model neuron outputs from a random network driven by time-varying external inputs (gray, network schematized in (A)) with no training ($p$ = 0). For this example, $pVar$ = 0.2%. Right panel shows the same PPC neurons as in the left panel in red, along with 4 model neuron outputs from a PINned network with $p$ = 12% plastic synapses. For this example, $pVar$ = 85%.

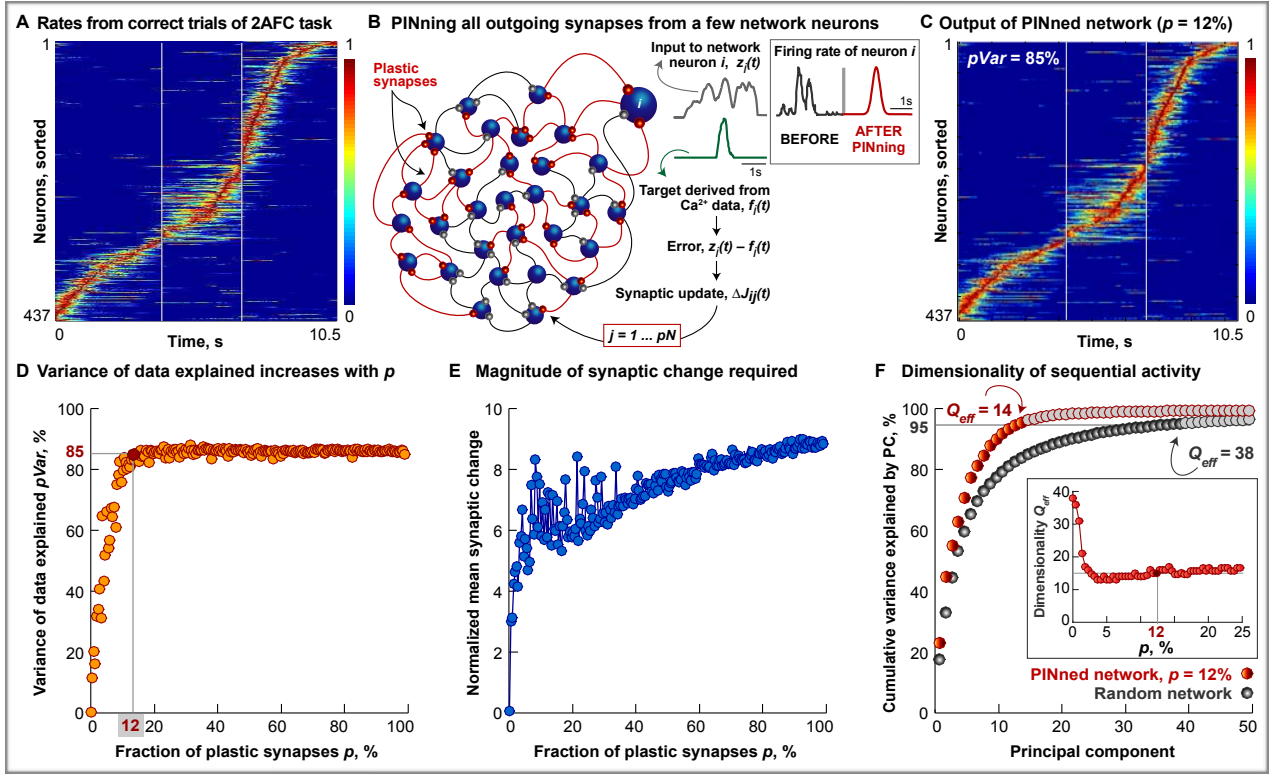

**Figure 2: Partial In-Network Training (PINning) Matches PPC-Like Sequences**

A. Identical to Figure 1C.

B. Schematic of the activity-based modification scheme we call Partial in-Network Training or PINning is shown here. Only the synaptic weights carrying the recurrent inputs from a small and randomly selected subset, controlled by the fraction $p$ of the $N$ = 437 firing rate neurons (blue) in the network are modified (plastic synapses depicted in orange) at every time step by an amount proportional to the difference between the input to the respective neuron at that time step ($z_i(t)$, plotted in gray), and a target waveform ($f_i(t)$, plotted in green), the presynaptic firing rate, $r_j$, and the inverse cross-correlation matrix of the firing rates of the PINned neurons (denoted by the matrix **P**, Experimental Procedures 4). Each neuron in the PINned network fires a bump like the one in the red trace with its peak at a time point staggered relative to the other neurons in the network, and together, the population spans the entire duration of the task. Here, the target functions we use for PINning are extracted from the firing rates shown in (A).

C. Normalized activity from the network with $p$ = 12% plastic synapses (also indicated by the red circle in (D)). The activity of the PINned network is temporally constrained, has a relatively small amount of extra-sequential activity ($bVar$ = 40% for (C) and (A)), and shows an excellent match with data ($pVar$ = 85%).

D. Effect of increasing the PINning fraction, $p$, in a network producing a single PPC-like sequence like (A), is shown here. $pVar$ increases from 0 for random networks with no plastic connections (i.e., $p$ = 0) to $pVar$ = 50% for $p$ = 8% (not shown), and asymptotes at $pVar$ = 85% as $p \geq$ 12% (highlighted by the red circle and outputs shown in (C)).

E. The total magnitude of synaptic change to the connectivity matrix as a function of $p$ is shown here (computed as described in Experimental Procedures 8 for the matrix, $J_{PINned,\ 12\%}$). The normalized mean synaptic change grows from a factor of ~7 for sparsely PINned networks ($p$ = 12%) to ~9 for fully PINned networks ($p$ = 100%) producing the PPC-like sequence. This means that although the individual synapses change more in small-$p$ networks, the total amount of change across the synaptic connectivity matrix is smaller.

F. Dimensionality of the sequential activity is computed (Experimental Procedures 5) by plotting the cumulative variance explained by the different principal components (PCs) of the 437-neuron PINned network generating the PPC-like sequence (orange circles) with $p$ = 12% and $pVar$ = 85%, relative to those of the random network ($p$ = 0, gray circles). Of the 437 possible PCs that can capture the total variability in the activity of this 437-neuron network, 14 PCs account for over 95% of the variance when $p$ = 12%. This effective dimensionality, denoted by $Q_{eff}$, is smaller than the 38 accounting for >95% of the variability of the random untrained network. **Inset** shows this effective dimensionality, $Q_{eff}$ (depicted in red circles), of the manifold of the overall activity of the network, as $p$ increases. $Q_{eff}$ drops from about 38 dimensions for $p$ = 0 to about 14 dimensions by $p \geq$ 12%, indicating therefore that even a fully PINned sequence network does not construct a low-dimensional dynamical solution. In comparison, $Q_{eff}$ of the data in Figure 1D is 24.

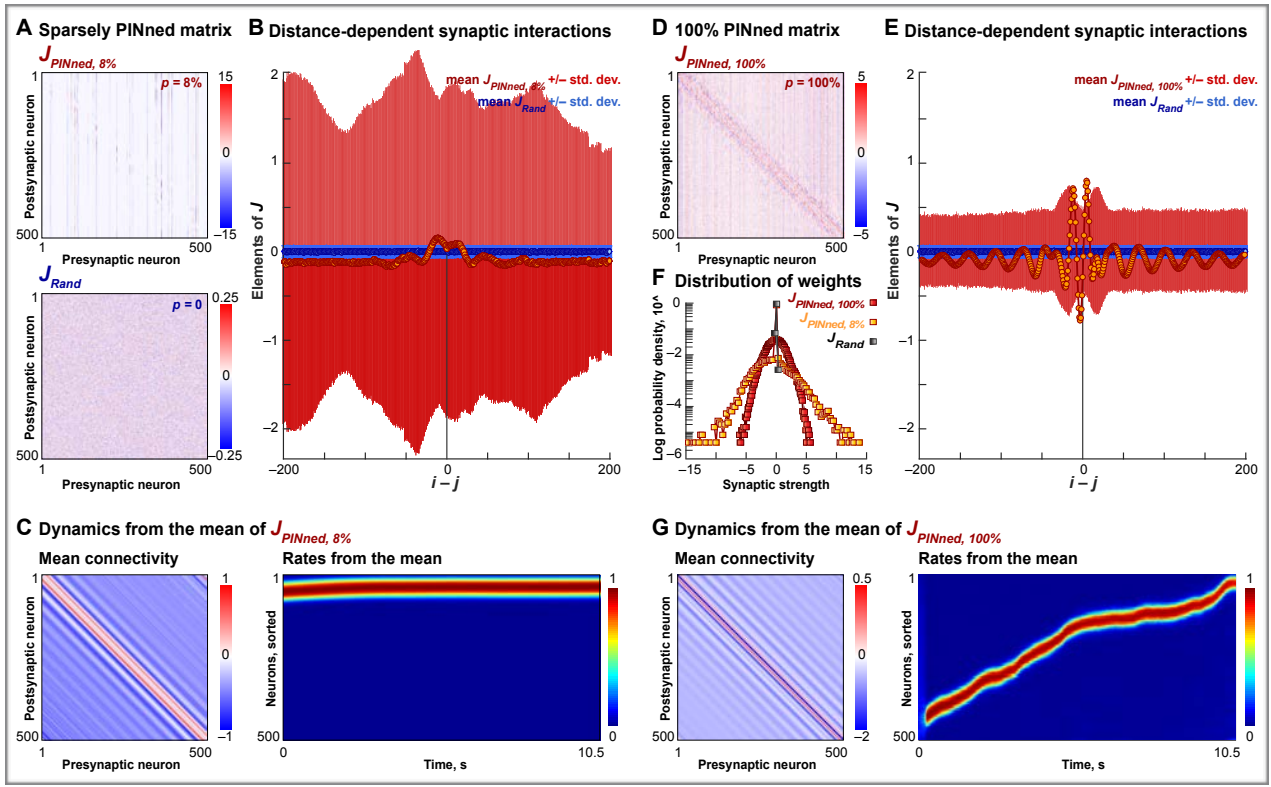

**Figure 3: Properties Of PINned Connectivity Matrices**

A. Synaptic connectivity matrix of a 500-neuron network with $p$ = 8% that produces an idealized sequence with $pVar$ = 92% (highlighted by the red circle in Supplemental Figure 4B), denoted by $J_{PINned, 8\%}$, and that of the randomly initialized network with $p$ = 0, denoted by $J_{Rand}$, are shown here. Colorbars on the panels indicate, in pseudocolor, the magnitudes of synaptic strengths after PINning.

B. Influence of neurons away from the sequentially active neurons is estimated by computing the mean (circles) and the standard deviation (lines) of the elements of $J_{Rand}$ (in blue) and $J_{PINned, 8\%}$ (in orange) in successive off-diagonal "stripes" away from the principal diagonal (as described in Experimental Procedures 11). These quantities are plotted as a function of the "inter-neuron distance", $i - j$. In units of $i - j$, 0 corresponds to the principal diagonal or self-interactions, and the positive and the negative terms are the successive interaction magnitudes of neurons a distance $i - j$ away from the primary sequential neurons. Fluctuations around the mean interaction are much larger and much more structured for $J_{PINned, 8\%}$ (red lines) compared to those for $J_{Rand}$ (blue lines).

C. Dynamics from the band-averages of $J_{PINned, 8\%}$ are shown here. Left panel is a synthetic matrix generated by replacing the elements of $J_{PINned, 8\%}$ by their respective means (orange circles in (B)). The normalized activity from a network with this synthetic connectivity is shown on the right. Although there is a localized "bump" of excitation around $i - j = 0$ and long range inhibition, these features are not responsible for sequences, but rather lead to a fixed point.

D. Same as panel (A), except for the synaptic connectivity matrix from a fully PINned ($p$ = 100%) network, denoted by $J_{PINned, 100\%}$ and shown here for comparison purposes. The initial random network with $p = 0$, $J_{Rand}$, is identical to the one in (A) and is omitted here.

E. Same as (B), except comparing $J_{PINned, 100\%}$ and $J_{Rand}$. Band-averages (orange circles) are bigger and more asymmetric compared to those for $J_{PINned, 8\%}$. Notably, these band-averages are also negative for $i - j = 0$ and in the neighborhood of 0. Fluctuations around the band-averages (red lines) for $J_{PINned, 100\%}$ are smaller than those for $J_{PINned, 8\%}$.

F. Log of the probability density of the elements of $J_{Rand}$ (gray squares), $J_{PINned, 8\%}$ (yellow squares) and for comparison, $J_{PINned, 100\%}$ (red squares) are shown here. $J_{Rand}$ is normally distributed with zero mean, variance = 0.005, and zero skewness and zero kurtosis. $J_{PINned, 8\%}$ is skewed toward negative or inhibitory weights (mean = –0.1, variance = 2.2, and skewness = –2) with heavy tails (kurtosis = 30). $J_{PINned, 100\%}$ has mean = 0, variance = 0.7, skewness = –0.02, and kurtosis = 0.2.

G. Same as panel (C), except showing the firing rates from the mean $J_{PINned, 100\%}$. In contrast with $J_{PINned, 8\%}$, the band averages of the fully PINned network can be sufficient to evoke a "Gaussian bump" that is qualitatively similar to the moving bump of activity evoked by a ring attractor model, since the movement of the bump is driven by the asymmetry in the mean connectivity.

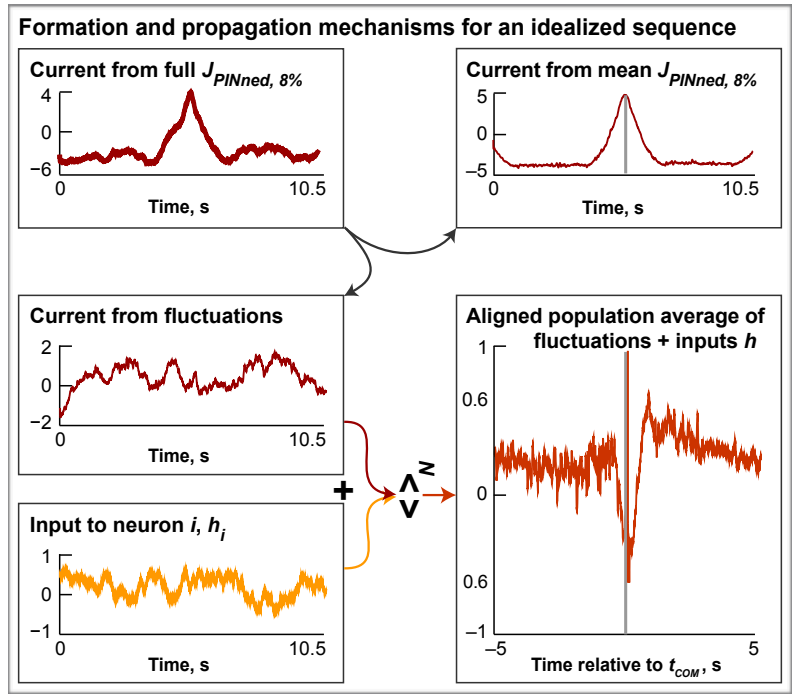

**Figure 4: Mechanism For Formation And Propagation Of Sequence**

The currents from the matrix $J_{PINned, 8\%}$ (red trace in the top left panel) and from its components are examined to uncover the mechanism for the formation and the propagation of a single idealized sequence. The band-averages of $J_{PINned, 8\%}$ cause the bump to form, as shown in the plot of the current from mean $J_{PINned, 8\%}$ to one neuron in the network (red trace in the top panel on the right, see also Figure 3C). On the other hand, the cooperation of the fluctuations around the means (whose currents are plotted in red in the middle panel on the left) with the external inputs (in yellow in the bottom left) causes the bump to move. We demonstrate this by considering the currents from the fluctuations around mean $J_{PINned, 8\%}$ for all the neurons in the network combined with the currents from the external inputs. The summed currents are realigned to the $t_{COM}$ of the bump (see Experimental Procedures 6 for a similar realignment) and then averaged over neurons. The resulting curve, an aligned population average of the sum of the fluctuations and external inputs to the network, is plotted in the bottom panel on the right, and reveals the asymmetry that is responsible for the movement of the bump across the network.

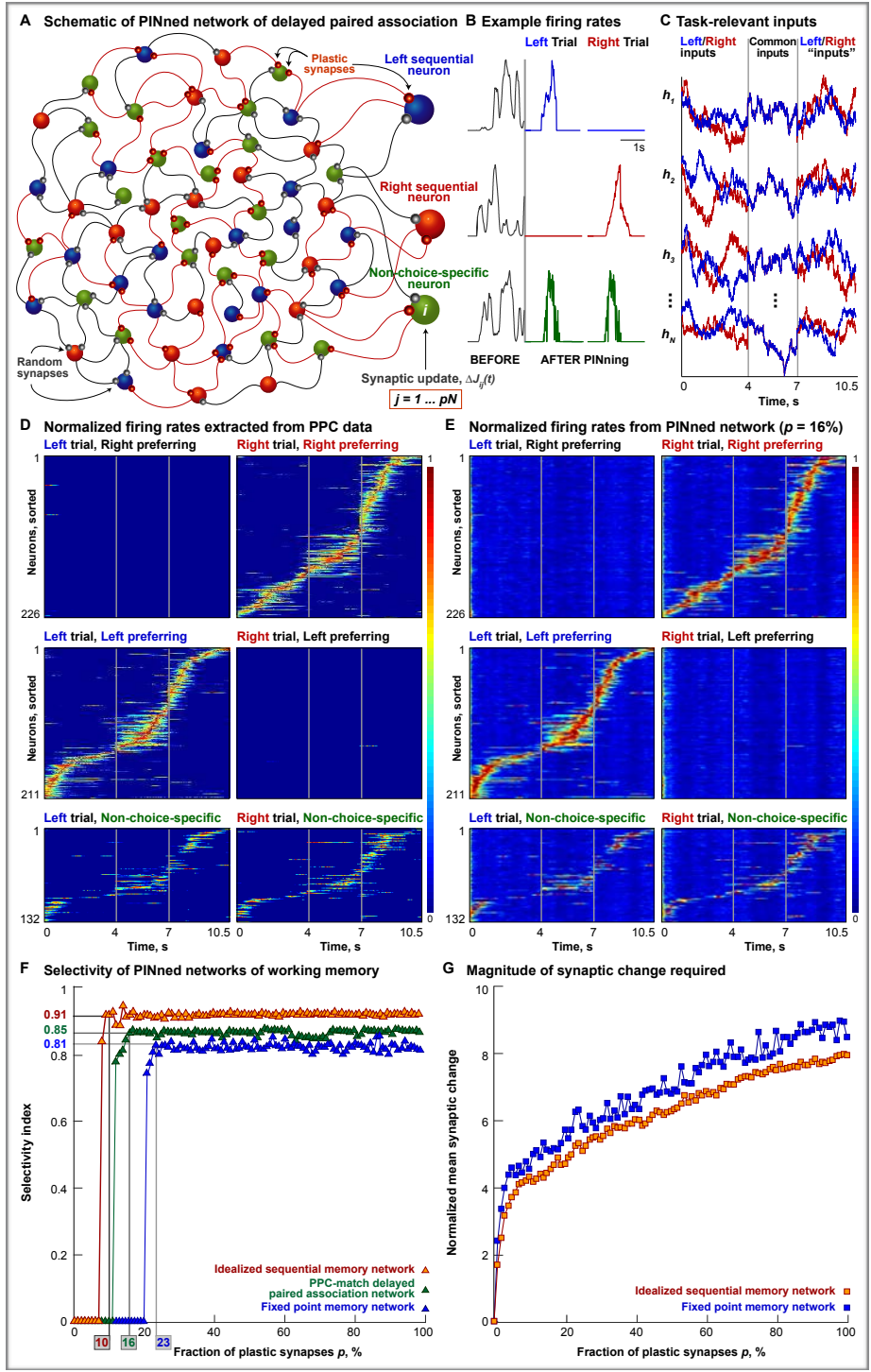

**Figure 5: Delayed Paired Association In PINned Networks Of Working Memory**

A. Schematic of a sparsely trained network constructed by PINning to implement a delayed paired association task, a two alternative forced-choice task [Harvey, Coen & Tank, 2012]. Only the outgoing synapses (plastic synapses are depicted in orange) from small subsets of randomly interconnected (random synapses are depicted in gray) network neurons are plastic, using as targets, the firing rates extracted from $Ca^{2+}$ imaging data from left preferring PPC cells (schematized in blue), right preferring cells (schematized in red) and cells with no choice preference (in green, called non-choice-specific neurons). As schematized in Figure 2B (see also Experimental Procedures 4), the learning rule is applied only to $p\%$ of the synapses in the network.

B. Example single neuron firing rates, normalized to the maximum, before (gray) and after PINning (blue trace for left preferring, red trace for right preferring and green trace for non-choice-specific neurons). After PINning, left preferring neurons (blue trace) are active only during a left trial and are silent during a right trial, right preferring neurons (red trace) are active only during a right trial, and non-choice-specific neurons (green trace) are active in trials of both types.

C. A few example task-specific inputs ($h_i$ for $i = 1, 2, 3, …, N$) are shown here. Each network neuron gets a different one of these irregular, spatially delocalized, filtered white-noise inputs (Experimental Procedures 2), but receives the same one on every simulated trial of the task. Inputs for the left trial are plotted in blue and those for the right trial are in red. During the first 4s of each 10.5s-long trial of the task, the cue period, each network neuron receives a different input, depending on whether that specific trial is a left trial or a right trial. During the middle 3s, the delay period, left- and right-specific inputs coalesce into a single time-varying pattern to simulate the fact that during the real delay period in the experimental task, visual inputs seen by the mouse are not choice-specific. During the final 3.5s of the trial, the turn period, left and right specific "inputs" are again separate and different, to simulate the distinct experiences the mouse might have while executing a left turn versus a right turn.

D. Normalized firing rates extracted from trial-averaged $Ca^{2+}$ imaging data collected in the PPC during a 2AFC task are shown here. Spike trains are extracted by deconvolution (Experimental Procedures 3) from mean calcium fluorescence traces for the 437 choice-specific and 132 non-choice-specific, task-modulated cells (one cell per row) imaged on preferred and opposite trials [Harvey, Coen & Tank, 2012]. These firing rates are used to extract the target functions for PINning (schematized in (A), Experimental Procedures 3 and 4). Traces are normalized to the peak of the mean firing rate of each neuron on preferred trials and sorted by the time of center-of-mass ($t_{COM}$). Vertical gray lines indicate the time points corresponding to the different epochs of the task – the cue period ending at 4s, the delay period ending at 7s, and the turn period concluding at the end of the trial, at 10.5s.

E. The outputs of the 569-neuron recurrent network with $p$ = 16% plastic synapses, sorted by $t_{COM}$ and normalized by the peak of the output of each neuron, showing a match with the experimental data (D). For this network, *pVar* = 85% (Experimental Procedures 7).

F. Selectivity index, shown here, is computed as the ratio of the difference and the sum of the mean activities of preferred neurons at the 10s time-point during preferred trials and the mean activities of preferred neurons during opposite trials (Experimental Procedures 9). Task performance of 3 different PINned networks of working memory as a function of the fraction of plastic synapses, $p$ – an idealized sequential memory model (orange triangles, $N$ = 500), a model that exhibits PPC-like dynamics (green triangles, $N$ = 569, using rates from (D)) and a fixed point memory network (blue triangles, N = 500, see also Supplemental Figure 5) – are shown here. The network that exhibits long-duration population dynamics and memory activity through idealized sequences (orange triangles) has selectivity = 0.91, when $p$ = 10% of its synapses are plastic; the PPC-like network needs $p$ = 16% plastic synapses for selectivity = 0.85, and the fixed point network needs $p$ = 23% of its synapses to be plastic for selectivity = 0.81.

G. Magnitude of synaptic change (computed as for Figure 2E, Experimental Procedures 8) for the 3 networks shown in (F) here. The idealized sequential memory network (orange squares) and the fixed point memory network (blue squares) require comparable amounts of mean synaptic change to execute the DPA task, growing from a factor of ~3 to ~9 for $p$ values ranging from 10% through 100%. This indicates that across the connectivity matrix, the total amount of synaptic change is smaller, even though individual synapses change more in sparsely PINned (small $p$) networks. The PPC-like memory network is omitted for clarity.

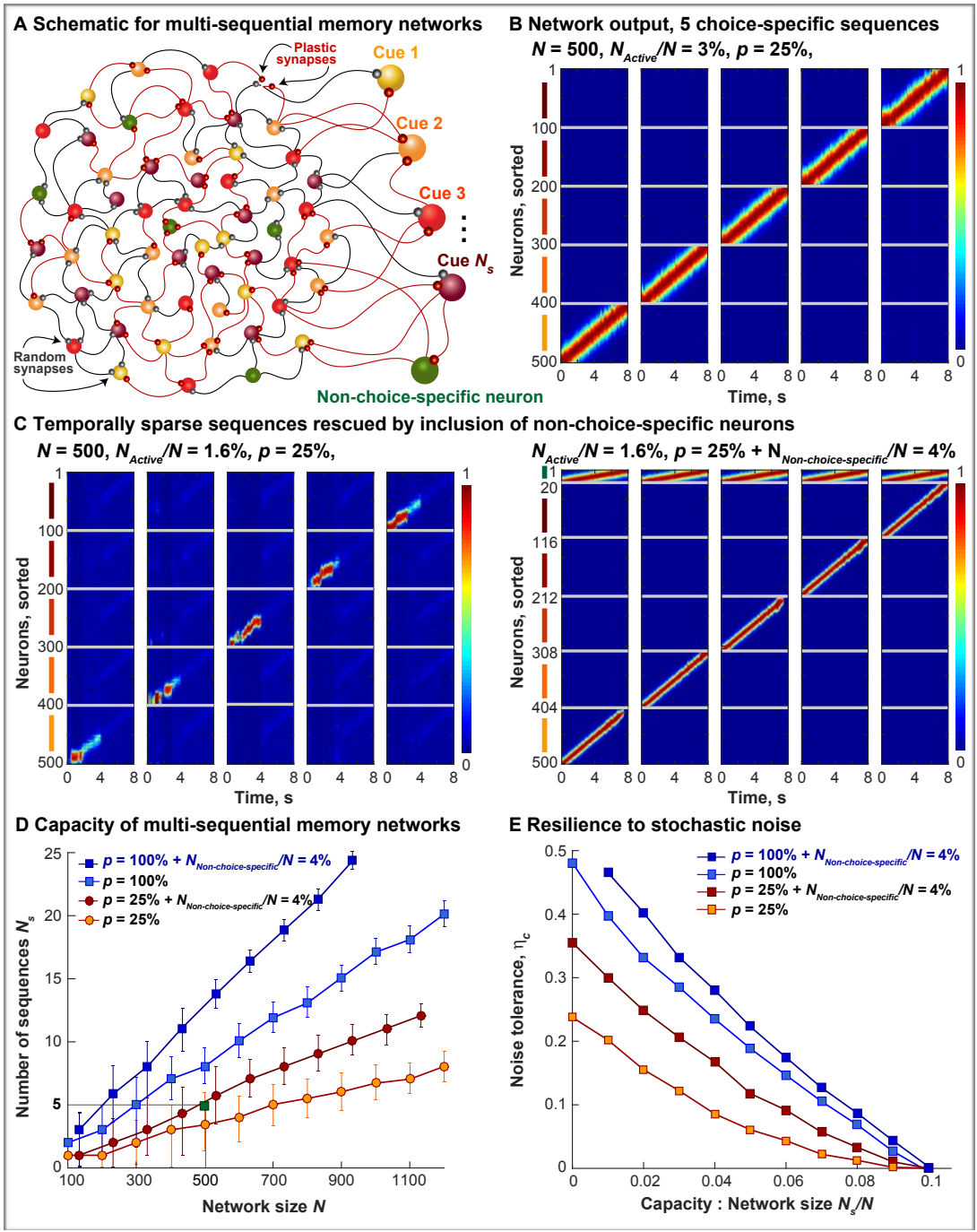

**Figure 6: Capacity Of Multi-Sequential Memory Networks**

A. Schematic of a multi-sequential memory network of rate-based neurons ($N = 500$). $N/N_s$ neurons are assigned to each one of the multiple sequences that we want the network to simultaneously produce. Here, $N_s$ is the network capacity, which is equal to the number of types of "trials", as well as the number of "cue preferences". As described in the text, only $p\%$ of the synapses in the network are plastic. The target functions are identical to those used in Supplemental Figure 4A, however, their widths, denoted by the ratio, $N_{Active}/N$, can vary as a task parameter that controls how many network neurons are active at any instant, the temporal sparseness of the sequence (Experimental Procedures 10).

B. The normalized firing rates of the 500-neuron network with $p$ = 25% and $N_{Active}/N$ = 3% are shown here. The memory task (the cue periods and the delay periods only) is correctly executed through 5 choice-specific sequences; during the delay period (4–8s), neurons fire in a sequence only on trials of the same type as their cue preference and are silent during other types of trials.

C. Left panel shows the same network as (B) failing to perform the task correctly when the widths of the targets was halved ($N_{Active}/N$ = 1.6%), sparsifying the sequences in time. While there is sequential activation, the memory of the cue identity is not maintained during the delay period. Right panel shows the result of including a small number of non-choice-specific neurons ($N_{Non\text{-}choice\text{-}specific}/N$ = 4%) that fire in the same order in trials of all 5 types. Including non-choice-specific neurons restores the memory of cue identity during the delay period, without requiring an increase in $p$.

D. Capacity of multi-sequential memory networks, $N_s$, as a function of network size, $N$, for different values of $p$ is shown here. Mean values are indicated by orange circles for $p$ = 25% PINned networks, red circles for $p$ = 25% + $N_{Non\text{-}choice\text{-}specific}/N$ = 4%, light blue squares for $p$ = 100% and dark blue squares for $p$ = 100% + $N_{Non\text{-}choice\text{-}specific}/N$ = 4% neurons. PINned networks containing additional non-choice-specific neurons have temporally sparse target functions with $N_{Active}/N$ = 1.6%, the rest have $N_{Active}/N$ = 3%. Error bars are calculated over 5 different random instantiations each and decrease under the following conditions: as network size is increased, as fraction of plastic synapses $p$ is increased, and moderately with the inclusion of non-choice-specific neurons. The green square highlights the 500-neuron network whose normalized outputs are plotted in (B). The maximum sequence-carrying capacity, $N_s$, is directly proportional to $pN$ and inversely proportional to $N_{Active}/N$.

E. Resilience of different multi-sequential networks is computed as the critical amount of stochastic noise (denoted here by $\eta_c$, Experimental Procedures 2) tolerated before the memory task fails, is shown here. Tolerance, plotted here as a function of the ratio of multi-sequential capacity and network size, $N_s/N$, decreases as $p$ is lowered and as capacity increases, although it falls slower with the inclusion of non-choice-specific neurons. Only mean values are shown here for clarity.

**SUPPLEMENTAL DATA**

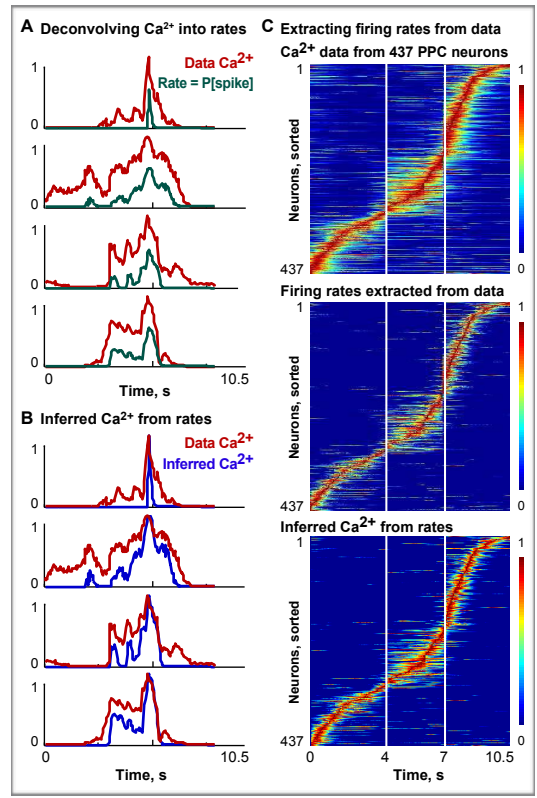

**Figure S1: Extracting Target Functions From Data By Deconvolution**

A. The use of deconvolution algorithms for the extraction of firing rates is illustrated here. A few example PPC neurons showing the $Ca^{2+}$ fluorescence signals in red (replotted from [Harvey, Coen & Tank, 2012]) and their extracted firing rates (normalized to a maximum of 1) in green. These rates were obtained using both methods described in Experimental Procedures 3. The implementation of the Bayesian algorithm (method #2 from Experimental Procedures 3) yields spike times along with the statistical confidence of a spike arriving at that particular time point (P[spike]). However, since the frame rate of the imaging experiments in [Harvey, Coen & Tank, 2012] was relatively slow (64ms per frame), we used these "probabilistic" spike trains as a normalized firing rate estimate. To verify, we first deconvolved single trial $Ca^{2+}$ data, and then performed an average over all the single trial P[spike] estimates to obtain the smooth firing rates for each of the 437 sequential neurons that we show in the middle panel of (C).

B. To see how accurate the firing rate estimates extracted from the data were, we re-convolved the extracted firing rates of the example units in (A) through a difference of exponentials with a rise time of 52ms and decay time of 384ms. The PPC $Ca^{2+}$ data is plotted in red and the inferred $Ca^{2+}$ is plotted in blue here.

C. Top panel shows the original $Ca^{2+}$ fluorescence signals from the PPC (similar to Figure 2A of main text, adapted from Figure 2c in [Harvey, Coen & Tank, 2012]). Middle panel shows the firing rates (from 0 to a maximum of 6Hz, normalized by the peak to 1) extracted from these data by using deconvolution methods. Bottom panel shows the inferred $Ca^{2+}$ signals obtained by convolving these extracted firing rates by a $Ca^{2+}$ impulse response function as explained for (B) (see also Experimental Procedures 3).

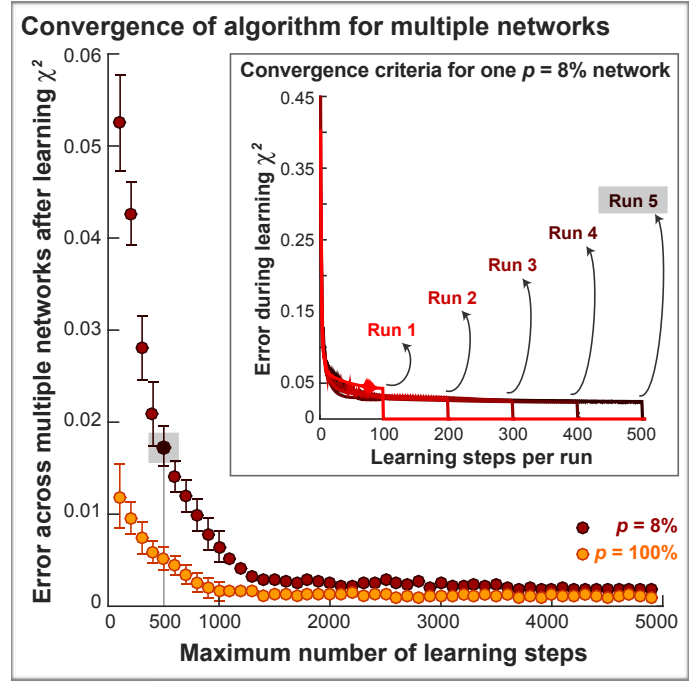

**Figure S2: Convergence Of PINning Algorithm**

The convergence of our PINning algorithm (Experimental Procedures 4) is shown here for multiple sequential networks generating sequential outputs (similar to Supplemental Figure 4). We plot the $\chi^2$ error between the target functions and the outputs of several PINned networks as a function of the number of learning steps for two values of $p$ – $p$ = 8% plastic synapses (mean values in the red circles, means computed over 5 instantiations each) and $p$ = 100% plastic synapses (mean values in the yellow circles, means computed over 5 instantiations each) for comparison purposes. When the $\chi^2$ error drops below 0.02, which for both sparsely and fully PINned networks occurs before the 500th learning step, we terminate the learning and simulate the network with the PINned connectivity matrix (denoted as $J_{PINned, p\%}$ in general) for an additional 50 steps before the program graphs the network outputs (firing rates, inferred calcium to compare with data, statistics of $J_{PINned, p\%}$, etc.). The point highlighted in the gray square corresponds to a PINned network with $p$ = 8% that ran for 500 learning steps, at the end of which, the $\chi^2$ error was 0.018. Additionally, this network had a *pVar* (Experimental Procedures 7) of 92%. **Inset** shows the speed of convergence for runs of different lengths for the $p$ = 8% PINned network. Run #5 corresponds to the example network highlighted in the gray square.

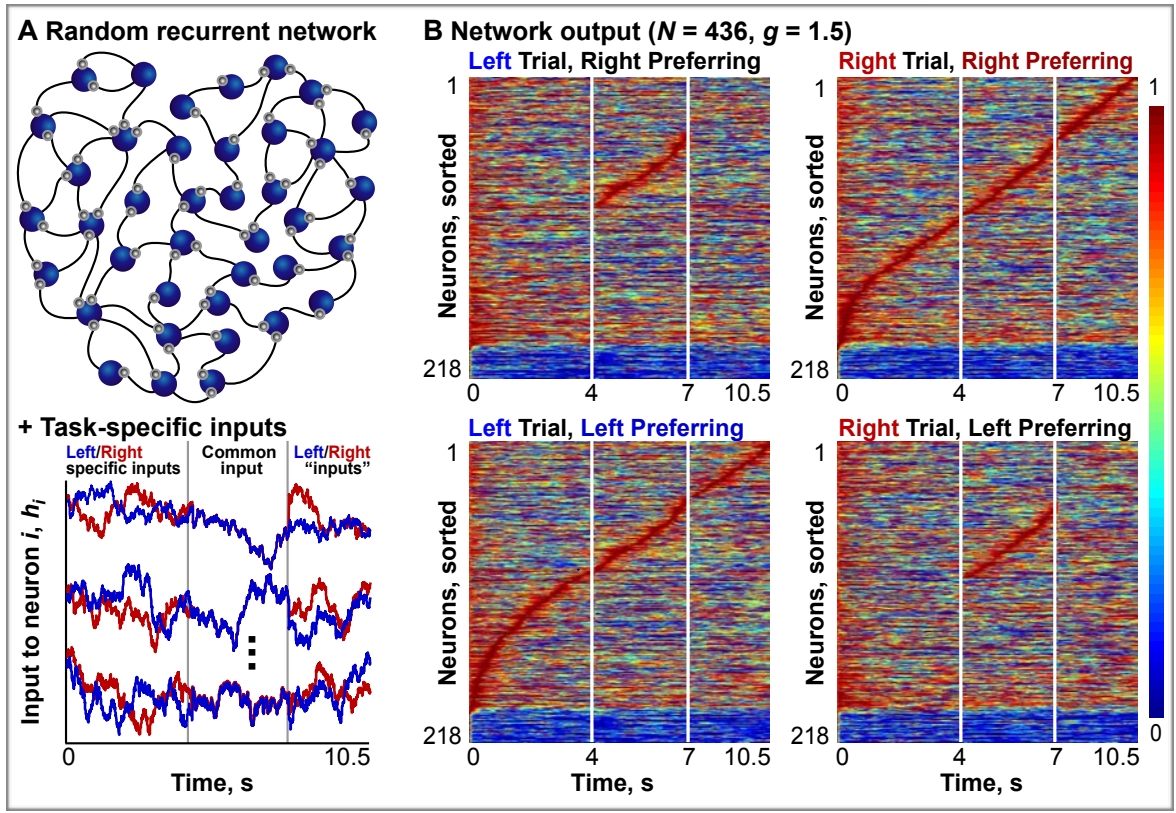

**Figure S3: No Delay Period Memory In Random Network Driven By Choice-Specific Inputs**

A. Schematic of a 436-neuron random recurrent network operating in a spontaneously active regime ($g = 1.5$, top panel) driven by filtered white noise inputs (Experimental Procedures 1 and 2) for a two alternative forced-choice task (bottom panel, same as Figure 5C). We assign 218 of the neurons in this network to be left preferring and the other 218 to be right preferring.

B. Normalized firing rates from the 436 random network neurons are sorted by their $t_{COM}$ and shown here. Both left preferring and right preferring neurons are active during the delay periods of both types of trials (between the 4s and 7s time points). The memory of the identity of the cue disappears during the delay period in this network because the inputs coalesce between 4–7s and become cue-invariant.

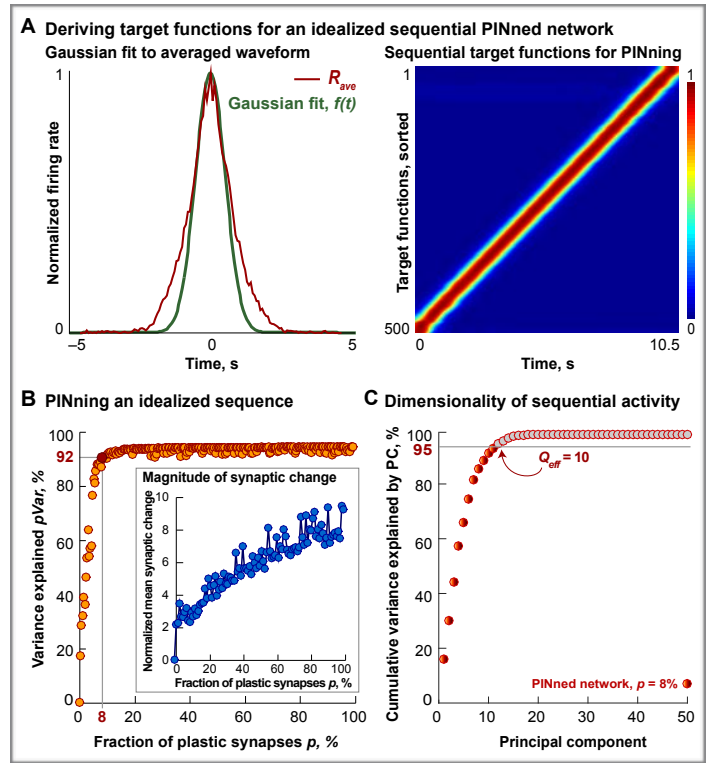

**Figure S4: Idealized Sequence-Generating PINned Network**

A. Left panel shows a Gaussian with mean = 0 and variance = 0.3, denoted by *f(t)* (green trace), that best fits the neuron-averaged waveform (red trace, $R_{ave}$, identical to Figure 1D). This waveform *f(t)* is used to generate the target functions (right panel) for a network of 500 rate-based model neurons PINned to produce an idealized sequence of population activity (*bVar* = 100%).

B. Effect of increasing the PINning fraction, *p*, in a network producing the single idealized sequence is shown here. *pVar* (Experimental Procedures 7) plotted as a function of *p*, increases from 0 for a random unmodified network (*p* = 0) network and plateaus at *pVar* = ~92% for and above *p* = 8%. The sequence-facilitating properties of the connectivity matrix in the *p* = 8% network highlighted in red are analyzed in Figure 4 of the main text. **Inset** shows the magnitude of synaptic change required to generate an idealized sequence as a function of *p*, computed as in Figure 2E (Experimental Procedures 8). The overall magnitude grows from a factor of ~3 for sparsely PINned networks (*p* = 8%) to between 8 and 9 for fully PINned networks (*p* = 100%) producing a single idealized sequence that match the targets in (A). As explained in the main text, although the individual synapses change more in sparsely PINned (small *p*) networks, the total amount of change across the synaptic connectivity matrix is smaller.

C. Dimensionality of sequential activity is computed (as in Figure 2F, Experimental Procedures 5) for the 500-neuron PINned network generating an idealized sequence (orange circles) with $p = 8\%$ and $pVar = 92\%$. Of the 500 possible PCs that can capture the total variability in the activity of this 500-neuron network, 10 PCs account for over 95% of the variance when $p = 8\%$, i.e., $Q_{eff} = 10$.

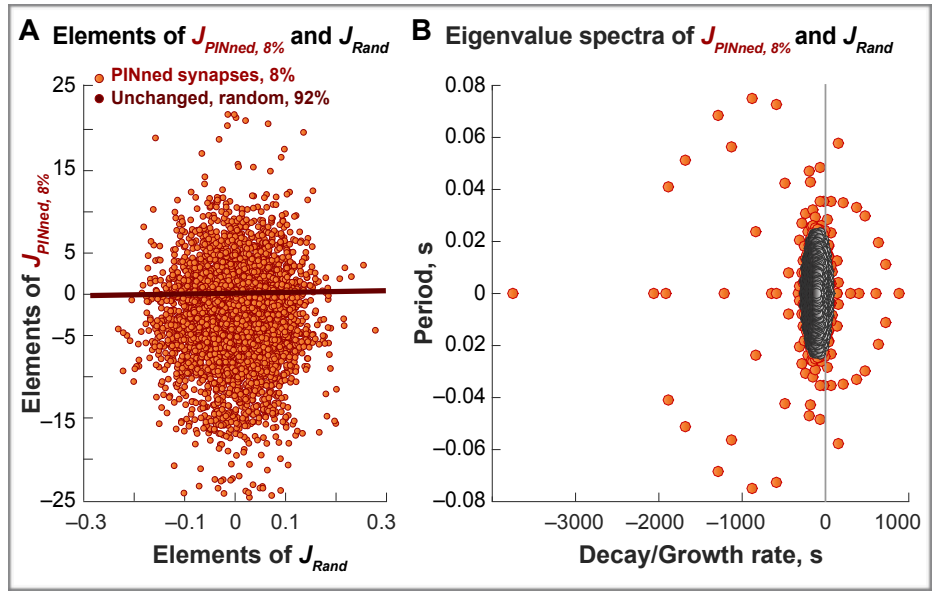

**Figure S5: Elements And Eigenvalues Of $J_{PINned, 8\%}$ Relative To $J_{Rand}$**

A. Elements of the matrices $J_{Rand}$ and $J_{PINned, 8\%}$ (red dots) are plotted against each other here. The largest changes in $J_{PINned, 8\%}$ are in the scattered 40,000 light red dots (8% of the total number of weights, $pN^2$) and there is a negative bias to their spread. Identity line is plotted in dark red.

B. The eigenvalue spectra of $J_{Rand} - I$ (gray circles) and $J_{PINned, 8\%} - I$ (orange circles) showing the time period (computed as $\mathrm{Re}(\lambda)/\tau$ where $\mathrm{Re}(\lambda)$ is the real part of the eigenvalues and $\tau$ is the time constant for network units) and decay/growth rates (computed as $\mathrm{Im}(\lambda)/2\pi\tau$ where $\mathrm{Im}(\lambda)$ is the imaginary part of the eigenvalues) corresponding to the different modes of the two networks. The gray line at 0 is the line of stability. As expected, the eigenvalues of $J_{Rand}$ lie uniformly within a circle, however, those of $J_{PINned, 8\%}$ are distributed non-uniformly with several large non-zero eigenmodes, some positive, others, more negative.

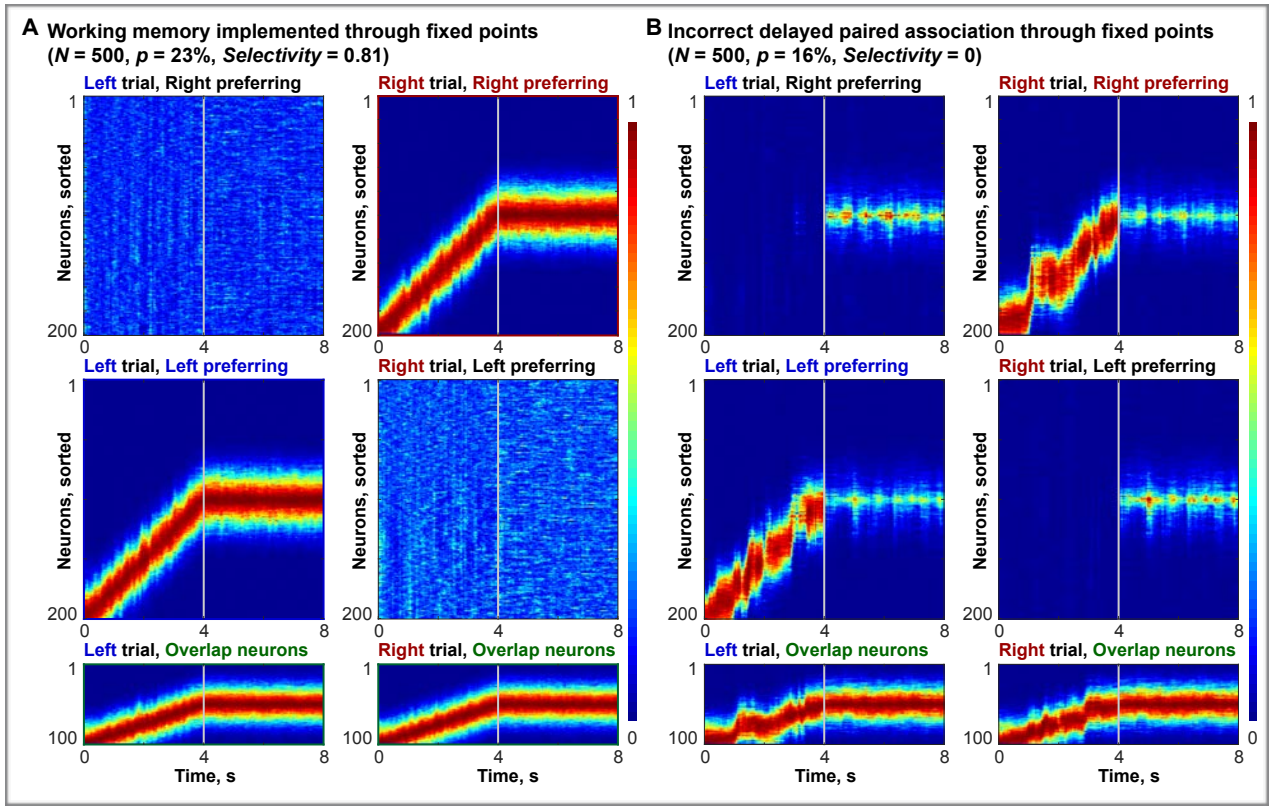

**Figure S6: Sequential Memory As Alternative Memory Mechanism To Fixed Points**

A. Working memory implemented through fixed points in a PINned network with $N = 500$ and $p = 23\%$ plastic synapses is shown here. This network has an asymptotic selectivity of 0.81, as shown in Figure 5F of the main text.

B. Inadequately PINned fixed point network ($p = 16\%$) fails to maintain cue memory during the delay period of the task (selectivity = 0).

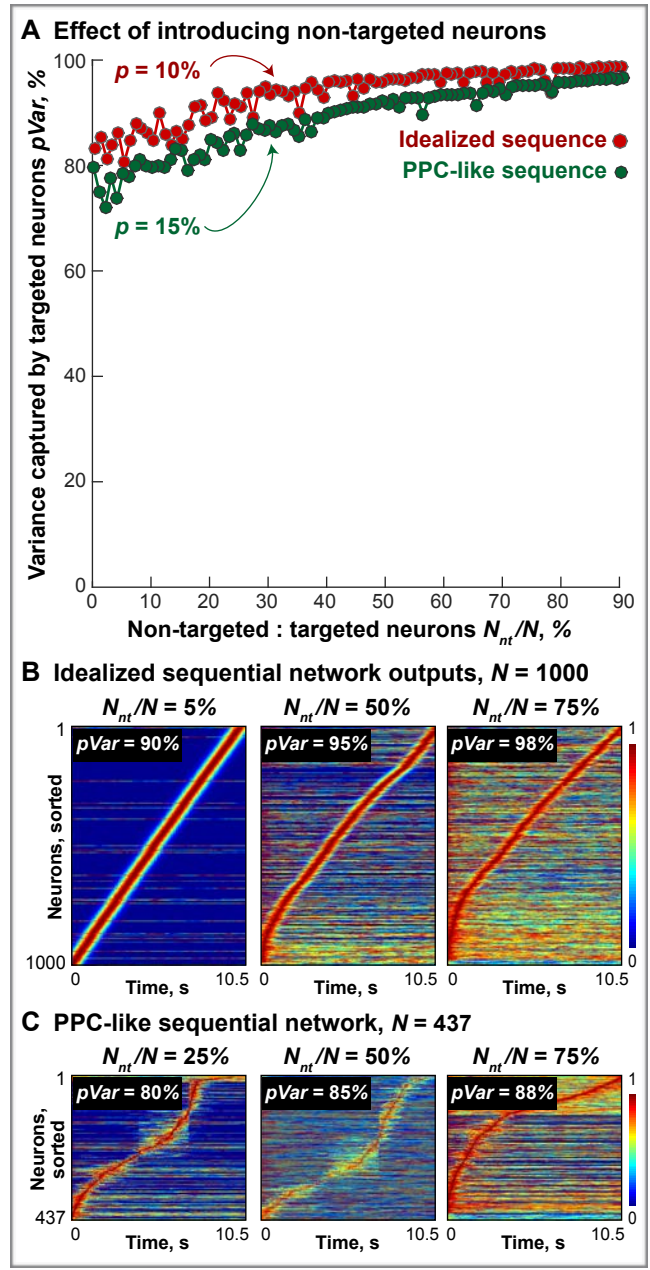

**Figure S7: Simulating Unobserved Neurons By Including Non-Targeted Neurons In PINned Networks**

A. Variance of the target functions captured by the outputs of the network neurons that have been PINned, *pVar* (evaluated as described in Experimental Procedures 7) is plotted here as a function of the ratio of non-targeted to targeted neurons in the network, denoted by $N_{nt}/N$. Overall, *pVar* does not decrease appreciably when the relative number of untrained neurons introduced into the PINned network is increased. Interestingly however, *pVar* improves slightly for sparsely PINned ($p$ = 10% networks in the red circles for an idealized sequence-generating network and $p$ = 15% networks in green for a PPC-like sequence) when untrained neurons are introduced. This improvement gets smaller as $p$ increases (not shown). The inclusion of non-targeted neurons in the networks constructed by PINning simulates the effect of unobserved but active neurons that may exist in the experimental data and might influence neural activity. It should be noted, however, that these additional neurons, however, might add irregularity to the sorted outputs of the full network (including both targeted and non-targeted neurons) and reduce the stereotypy of the overall outputs (indicated by a decrease in the *bVar* computed over the full network, not shown). This effect is independent of $p$.

B. Example network outputs are shown here for 3 values of $N_{nt}/N$. for a network generating an idealized sequence similar to Supplemental Figure 4. *pVar* = 90% for $N_{nt}/N$ = 5%, 95% for $N_{nt}/N$ = 50% and 98% for $N_{nt}/N$ = 75%. The sequences become noisier overall as more randomly fluctuating untrained neurons are introduced, but the percent variance of the targets captured by the PINned neurons remains largely unaffected, even showing a slight improvement.

C. Same as panel (B), except for a 437-neuron network constructed by PINning to generate a PPC-like sequence similar to Figure 2C with different fractions of neurons left untrained. *pVar* = 80% for $N_{nt}/N$ = 25%, 85% for $N_{nt}/N$ = 50% and 88% for $N_{nt}/N$ = 75%, confirming the same general trend as above.

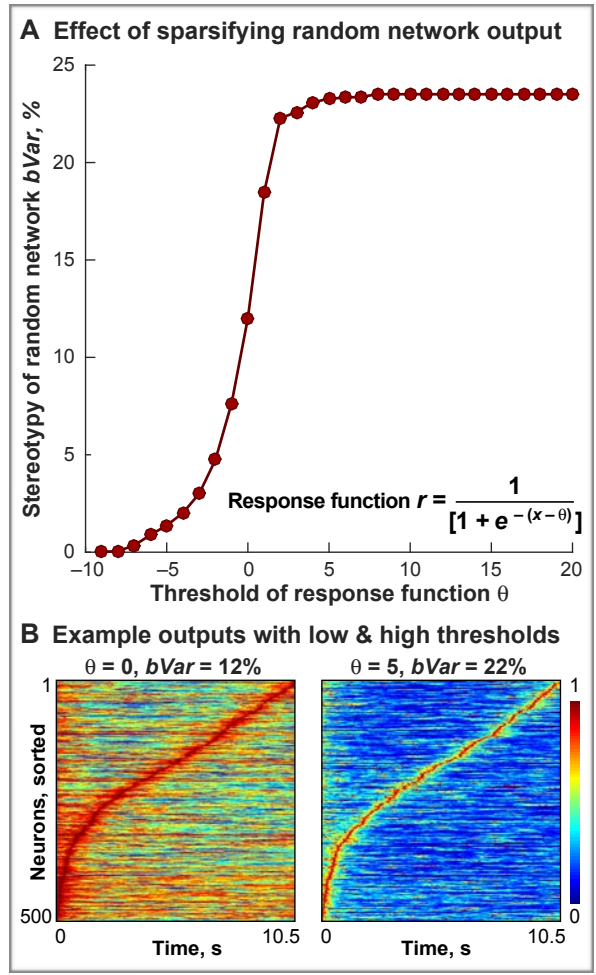

**Figure S8: Changing The Threshold Of The Response Function And Sparsifying The Random Network Output**

A. Effect of sparsifying the outputs of random networks of model neurons constructed with different thresholds, denoted by $\theta$, is shown here. Stereotypy of the sequences made by sorting the outputs of these random networks, *bVar*, increases from 0 to 12% for $\theta = 0$ networks used as initial configurations throughout the paper, and saturates at 22% for networks with threshold values of $\theta > 2$.

B. Example outputs from two random networks, one with $\theta = 0$ (left, with *bVar* = 12%, this is identical to Figure 1B) and one with $\theta = 5$ (right, with *bVar* = 22%) are shown here. Firing rates are normalized by the $t_{COM}$ and sorted to yield the sequences shown here.

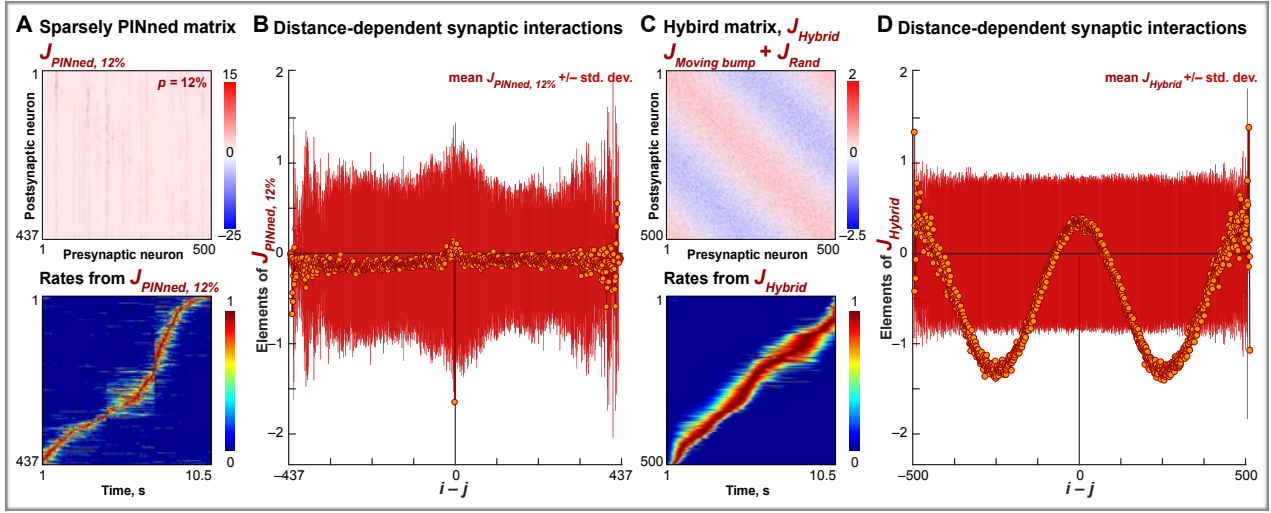

**Figure S9: Comparison Of Sparsely Pinned Matrix, $J_{Pinned,\ 12\%}$ And An Additive Hybrid Connectivity Matrix Of The Form, $J_{Hybrid} = J_{Moving\ Bump} + J_{Rand}$**

A. Synaptic connectivity matrix of a 437-neuron network with $p$ = 12% that produces a PPC-like sequence with $pVar$ = 85%, denoted by $J_{PINned,\ 12\%}$, and the firing rates obtained (identical to Figure 2C, also highlighted by the red circle in Figure 2D) are shown here.

B. Influence of neurons away from the sequentially active neurons is estimated by computing the mean (circles) and the standard deviation (lines) of the elements of $J_{Rand}$ (in blue) and $J_{PINned,\ 8\%}$ (in orange) in successive off-diagonal "stripes" away from the principal diagonal (as described in Experimental Procedures 11 and analogous to Figure 3 in the main text). These quantities are plotted as a function of the "inter-neuron distance", $i - j$. In units of $i - j$, 0 corresponds to the principal diagonal or self-interactions, and the positive and the negative terms are the successive interaction magnitudes of neurons a distance $i - j$ away from the primary sequential neurons.

C. Same as panel (A), except for the synaptic connectivity matrix from an additive hybrid of the form, $J_{Hybrid} = J_{Moving\ bump} + J_{Rand}$ is shown here, where, $J_{Rand.}$ is a random matrix similar to the one shown in the lower panel of Figure 3A and $J_{Moving\ bump}$ is the connectivity for a moving bump model [Yishai, Bar-Or & Sompolinsky, 1995]. The hybrid matrix contains a structured and a random part, and is constructed by the addition of a moving bump connectivity matrix [Yishai, Bar-Or & Sompolinsky. 1995],

$J_{\text{Moving bump}} = -J_0 + J_2[\cos(\phi_i - \phi_j)] + 0.06 \times [\sin(\phi_i - \phi_j)]$, $\phi = 0,...,\pi$) and a random matrix, $J_{Rand}$, similar to the one used to initialize PINning (lower panel of Figure 3A). Mathematically, the hybrid is of the form, $J_{\text{Hybrid}} = \frac{1}{N}[A_B J_{\text{Moving bump}}] + \frac{1}{\sqrt{N}}[A_R J_{\text{Rand}}]$, where $A_B$ is the relative amplitude of the structured or moving bump part, scaled by network size, $N$, and $A_R$ is the relative amplitude of the random part of the N-neuron hybrid network, scaled by $\sqrt{N}$. The lower panel shows the firing rates from the additive hybrid network whose connectivity is given by $J_{Hybrid}$. *pVar* for the output of this hybrid network is only 1%, however, its stereotypy, *bVar* = 92%.

D. Same as (B), except for $J_{Hybrid}$. Band-averages (orange circles) are bigger and more asymmetric compared to those for $J_{PINned, 12\%}$. Notably, these band-averages are positive for *i* – *j* = 0 and in the neighborhood of 0. Fluctuations around the band-averages (red lines) for $J_{Hybrid}$ are less structured than those for $J_{PINned, 12\%}$ and result from $J_{Rand}$.

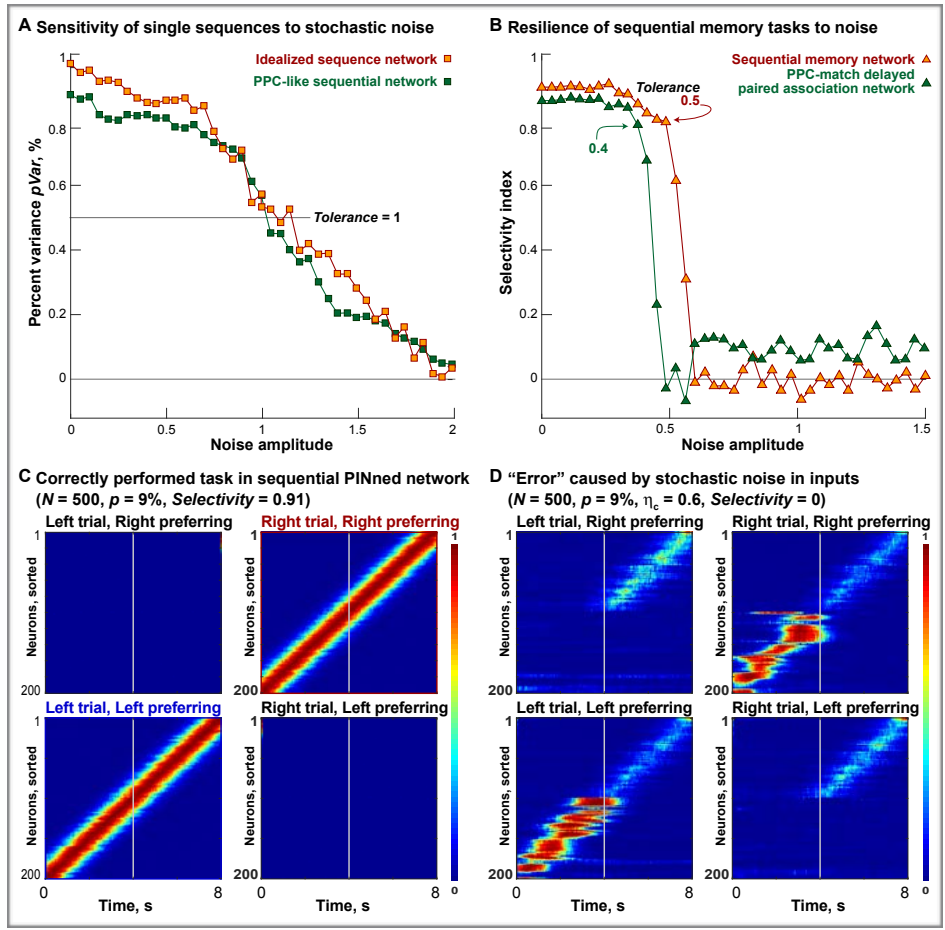

**Figure S10: Robustness Of PINned Networks To Stochastic Noise In Inputs**

A. Stochastic noise, $\eta$ is injected as an additional current to test whether, and how much, the neural sequences learned through PINning are stable against perturbations (as described in Experimental Procedures 2), and the results are shown here. Percent variance of the target functions explained by the network outputs, *pVar*, drops as amplitude of the injected noise is increased. The maximum tolerance is indicated by the gray line and for single sequences, we define it as the amplitude of noise at which *pVar* drops below 50% and denote it by $\eta_c$. For the idealized sequence and for the PPC-like sequence, $\eta_c = 1$.

B. Noise tolerance of memory networks that implement working memory through an idealized sequence (orange triangles) and through PPC-like sequences (green triangles, identical to the network shown in Figure 5) is plotted here. Selectivity of both networks drops at different values of $\eta$, indicating the maximum resilience or noise-tolerance of each, denoted by $\eta_c$ – the sequential memory network has a maximum tolerance of $\eta_c = 0.5$ while the PPC-match memory network has a maximum noise tolerance of $\eta_c = 0.4$.

C. Correctly performed delayed paired association task with delay period memory of the cue identity implemented through two idealized sequences (each similar to Supplemental Figure 4A) in a network of 500 neurons with *p* = 9% plastic synapses. For the outputs shown here, the turn period is omitted for clarity, and the delay period starts at the 5s time-point in the 10.5s-long task. The performance of this network, quantified by the selectivity index (Experimental Procedures 9) is 0.91.

D. The same network from panel (D) fails to perform the task (selectivity = 0) because the high levels of stochastic noise present in the inputs ($\eta = 0.6$, here) quenches delay period memory.

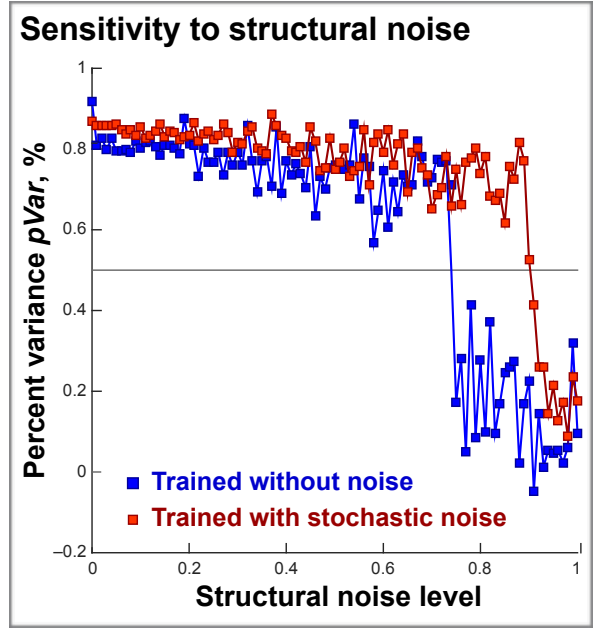

**Figure S11: Structural Noise Sensitivity Of PINned Networks**

Structural noise (described by $\text{level} \times \frac{\mathcal{N}(0,1)}{\sqrt{N}}$, where N = 500 here) is added to the connections in the sparsely PINned connectivity matrix, $J_{PINned,\ 8\%}$ to test whether, and by how much, the synaptic connections are finely tuned. *pVar*, computed as in Experimental Procedures 9, is plotted as a function of structural noise amplitude, for sequential networks obtained by PINning in the presence of stochastic noise in the inputs (red squares) and for networks trained without any noise (blue squares). Gray line is at *pVar* = 50%, the noise-free-PINned network has a tolerance (noise amplitude at which *pVar* drops below 50%) of 0.6, and the network trained with noise, 0.8. Training in the presence of stochastic noise therefore leads to slightly more robust networks, although the drop in *pVar* with the addition of structural noise does not fully recover with training noise.

**Supplemental Data S12: Cross-Validation Analysis**

We first divided the data from 436 neurons in Figure 1D into two separate "synthetic" sequences by assigning the even-numbered cells to one (let's call this Sequence A, containing 218 neurons) and the odd-numbered cells to another sequence (say, Sequence B, also with 218 neurons). Then we constructed a PINning-based network with 218 model neurons, exactly like we described in the main text, using data from Sequence A as target functions for PINning. Next, we computed the percent variance of the data in Sequence A and the data in Sequence B, denoted by $pVar_{Test\ A,\ Train\ A}$ and $pVar_{Test\ B,\ Train\ A}$, respectively, that are captured by the outputs of the PINned network (Experimental Procedures 7). Following a similar procedure, we also computed $pVar_{Test\ A,\ Seq\ B}$ and $pVar_{Test\ B,\ Seq\ B}$, after PINning a second network against target functions derived from Sequence B.

We obtained the following estimates for all fractions of plastic synapses, $p$ :

$$pVar_{Test\ A,\ Train\ A} = 91 \pm 2\% \qquad pVar_{Test\ B,\ Train\ A} = 45 \pm 2\%$$

$$pVar_{Test\ A,\ Train\ B} = 46 \pm 2\% \qquad pVar_{Test\ B,\ Train\ B} = 90 \pm 2\%$$

For comparison purposes, a random network such as the one in Figure 1B only captures a tiny amount of the variability of data (in this notation, $pVar_{Test\ Data,\ Random\ Network} = 0.2\%$, see also, right panel of Figure 1H). Additionally, the data from one set only accounts for 49% of the variance of the data of the other set, i.e., $pVar = 49\%$. Thus the model does almost as well as it possibly could.

**ACKNOWLEDGEMENTS**

The authors thank Larry Abbott for providing guidance and critiques throughout this project; Eftychios Pnevmatikakis and Liam Paninski for the deconvolution algorithm [Pnevmatikakis et al, 2014; Vogelstein et al, 2010]; and Dmitriy Aronov, Bill Bialek, Selmaan Chettih, Cristina Domnisoru, Tim Hanks, and Matthias Minderer for comments. This work was supported by the NIH (DWT: R01-MH083686; RC1-NS068148; 1U01NS090541-01 and CDH: R01-MH107620; R01-NS089521), a grant from the Simons Collaboration on the Global Brain (DWT), a Young Investigator Award from NARSAD/Brain & Behavior Foundation (KR), a fellowship from the Helen Hay Whitney Foundation (CDH), and a Burroughs Wellcome Fund Career Award at the Scientific Interface (CDH). CDH is a New York Stem Cell Foundation-Robertson Investigator and a Searle Scholar.